# Hybrid quantum network design against unauthorized secret-key generation, and its memory cost

Omer Sakarya,[1] Marek Winczewski,[2,3] Adam Rutkowski,[4] and Karol Horodecki[1,3]

[1]*Institute of Informatics, National Quantum Information Centre, Faculty of Mathematics, Physics and Informatics, University of Gdańsk, Wita Stwosza 57, 80-308 Gdańsk, Poland*
[2]*Institute of Theoretical Physics and Astrophysics, National Quantum Information Centre, Faculty of Mathematics, Physics and Informatics, University of Gdańsk, Wita Stwosza 57, 80-308 Gdańsk, Poland*
[3]*International Centre for Theory of Quantum Technologies, University of Gdańsk, Wita Stwosza 63, 80-308 Gdańsk, Poland*
[4]*Institute of Theoretical Physics and Astrophysics, University of Gdańsk, 80-308 Gdańsk, Poland*

A significant number of servers that constitute the Internet are to provide private data via private communication channels to mutually anonymous registered users. Such are the servers of banks, hospitals that provide cloud storage and many others. Replacing communication channels by maximally entangled states is a promising idea for the quantum-secured Internet (QI). While it is an important idea for large distances secure communication, for the case of the mentioned class of servers pure entanglement based solution is not only unnecessary but also opens a threat. A crack stimulating a node to generate secure connections via entanglement swapping between two hackers can cause uncontrolled consumption of resources. Turning into positive a recently proven no-go result by S. Bäuml *et al.* [Nat. Commun. **6**, 6908 (2015)], we propose a natural countermeasure against this threat. The solution bases on connections between hub-nodes and end-users realized with states that contain secure key but do not allow for swapping of this key. We then focus on the study of the quantum memory cost of such a scheme and prove a fundamental lower bound on its memory overhead. In particular, we show that to avoid the possibility of entanglement swapping, it is necessary to store at least twice as much memory than it is the case in standard quantum-repeater-based network design. For schemes employing either states with positive partial transposition that approximates certain privates states or private states hardly distinguishable from their attacked versions, we derive much tighter lower bounds on required memory. Our considerations yield upper bounds on a two-way repeater rate for states with positive partial transposition (PPT), which approximates strictly irreducible private states. As a byproduct, we provide a lower bound on the trace distance between PPT and private states, shown previously only for private bits.

## I. INTRODUCTION

The domain of quantum information processing, which shows how the rules of quantum mechanics can meet the needs of information society [1,2], has reached its maturity in recent years. We are about to enter the NISQ era of quantum computing with the noisy intermediate scale quantum (NISQ) devices ahead of us [3]. In parallel, a huge effort has been done towards building the quantum Internet (QI) [4–6], which is predicted to be built within several years [7]. It is viewed as a network of NISQ devices with their memory and the central processing unit (CPU), which exchange *qubits* rather than classical bits between each other.

The main welcome feature of the Qquantum Internet in comparison with the traditional Internet is its, speaking of theory, the *inherent security of sent signals*. The first-generation QI [5] bases on the quantum correlations called *entanglement* and its advantageous property of *transitivity*. In theory, a two otherwise disconnected nodes can obtain mutual unconditionally secure connection if only they share maximally entangled state (singlet) with a common node, via the *entanglement swapping* protocol [8,9]. Due to the high attenuation of quantum signals in optical fiber and impossibility of their amplification by cloning [10], the number of intermediate nodes which perform entanglement swapping (*quantum repeaters* [4]), needs to be large, and function in high coordination. Let us recall here that the quantum repeaters protect sent qubits against eavesdropping because entanglement swapping uses, in fact, quantum teleportation [9]. Indeed, quantum teleportation protocol allows for a transfer of data without any intermediate point in space-time, where it could be attacked.

While the QI is about to come, a number of serious attacks on the traditional Internet which is working already for about a halve a century is being more and more often reported. This happens in accordance with a growing interest in network cybersecurity. One of the simplest attacks on the network is the hijacking of a node, via a *malware*—a malicious piece of software which changes its functioning at a wish of a

---

hacker. Possible attacks on future quantum Internet has been recently considered [11,12]: a piece of software infects the CPU of a quantum device of the node of quantum repeater, leading, e.g., to local change of topology of the network. While proposals for overcoming the implications of such an attack are developed, we focus on a solution which to some extent, prevents it due to laws of physics.

*Hybrid quantum network.* As it is common in quantum information theory, a no-go (impossibility) in processing of quantum data can be exploited as its potential: quantum no-cloning led to the seminal ideas of quantum money and quantum cryptography protocols [1,2] while impossibility of prediction of measurement outcomes (attributing the so-called *hidden variable model*) led further to the device independent quantum security [13,14]. Our countermeasure to hijacking is also based on a recently found no-go, which can be stated as follows. *There exist quantum states which allow for point-to-point security of classical data against quantum adversary, and in spite of this fact can not be effectively used in quantum key repeaters* [15].

The above result shows that quantum security is not always *transitive*: for certain states (call them *nonrepeatable secure states* $\rho$), conversely to entanglement swapping, when $A$ has secure link (possessing $\rho$) with $B$ and $B$ with $C$, there is no possibility for $B$ to help $A$ and $C$, via a three-partite local quantum operations and classical communication (3-LOCC), to share a secure link, protected against $B$ as well. Certain bound entangled states [16] (from which no pure entanglement can be distilled by local operations and communication [17]) and highly noisy private states [18], has been recently shown, to fit the scheme in case of arbitrary 3-way and one-way classical communication (from the node $B$ to $A$ and $C$) respectively [15,19].

In this manuscript, we propose a general idea of *physical protection against malware* by presenting a flip side of the presented limitation on quantum repeaters. It amounts to deliberate use of the quantum states which disallow for repeating of secure key, in order to protect against any unauthorized network user who wants to perform it for his own purposes.

In the language of computer science, we propose an architecture and a model of the physical layer of the quantum network to exclude the possibility that its local topology is changed via attacking the network at the application layer.

We show that specially designed *hybrid quantum network*, i.e., based on both *repeaters* and special *relay stations* called here *hubs*, is more robust against special kind of attacks than original repeaters. We put forward a particular example of an attack and study properties of its countermeasure. To show the idea, we focus on a subnetwork of the hybrid quantum network, whose graph is a star, i.e., with a *central hub-node* and several ($\Delta$) connected *end-nodes* (see Fig. 1).

Our approach suits the scenario in which (1) the hub-node can be connected by a quantum repeater with other hub nodes; (2) only classical data need to be sent between the hub node and the end-nodes; (3) the distance between the hub-node and the end-nodes is maximally of metropolitan scale (up to the distance available for repeaterless quantum networks [20,21]); (4) only disconnected hub-nodes and their two adjacent end-nodes are attacked at a time; and (5) the attack is honest but curious: only functioning of the classical processor

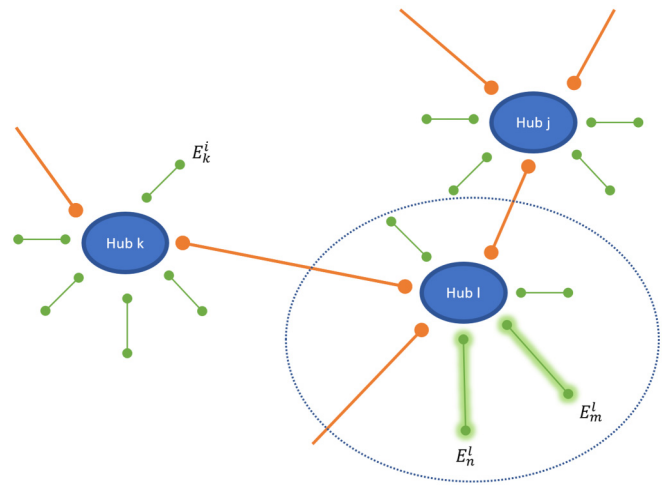

FIG. 1. Structure of the proposed hybrid network: thin green lines connect end-users with hubs; in these, only classical data can be transferred. Thick orange lines connect hubs being routing nodes of the network; these connections allow for passing a quantum state. Shaded lines connect two end-users that communicate securely with the hub node only classical data. Selected region denotes single hub of our interest.

is changed by malware, while classical data at the node remain unread.

The hybrid network is shown in Fig. 3. The above example fits the real use case, as in the network of the traditional Internet. Indeed in the Internet, there is quite a number of nodes representing servers that deliver certain utilities in the form of classical data, access to which is charged, and limited to a group of registered users. Moreover, the task of these nodes is not to connect the users, that are usually anonymous but to provide them an access to data via private link. Servers for online banking, access to medical data of a laboratory, online shops, and last but not least, providers of the data clouds form far from a complete list of examples of the latter. In some of these cases, the users are local so that the assumptions about the distance between end-nodes is satisfied. We focus on a star-shaped network with central hub connected to end-users. In this case, the data are generated in classical form. The distance between the users is usually not too big. It is also, needless to say, that security is vital since it is important for end-users, users of the hub, administrators or the owner of the server. We also focus on the case when two dishonest users of the network hijack a single node. Their task is to obtain a free secure connection. In other words the attack is a theft of processor time and power aimed at generation of secret-key. The main feature of our solution is that the *topology of the network* is naturally, physically protected against modification.

As it is usual, any good comes at a price. In the above case, the price will come out in the number of qubits needed to be stored (or processed) in quantum memory of a node. In the NISQ era, it is of prime importance to find how much of quantum memory is necessary in order to realize a given quantum network architecture. This is because, as of now, there is no technology to store coherently qubits for a long time. We therefore study lower bounds on the memory cost related to the hybrid quantum network and also show that this architecture can be realized with relatively modest quantum memory requirements. In principle, to form the links of a quantum network, one can take any state containing the private key. Useful from a cryptographic point of view are in that respect the so-called private states [18,22] or their approximate versions with positive partial transposition. The private states have directly accessible key via measurement on their subsystems called key parts. However, especially those having low repeatable key [19], have also the shielding systems (shields). A shield protects the key but costs quantum memory. As we will show, these states are not the only ones that have memory cost.

We therefore first provide lower bounds on the memory cost of our secure network scheme, which is related to the *density of the secure key in quantum states* - a natural quantity that was implicitly used in Refs. [15,22,23]. To our knowledge, this quantity has not been explicitly studied on its own so far. We introduce a *memory overhead* as a measure of the cost. For a *scheme* $S_\rho$ (that assures security of the hub node), its overhead is defined as

$$V(S_\rho) := M(\rho)(1 - \mathcal{D}(\rho)), \quad (1)$$

where $\mathcal{D}$ is the density of the key, i.e., ratio of the key to the dimension of the state, and $M(\rho)$ is the total memory of the scheme. This intuitive quantity is 0 for maximally entangled states, as their whole memory has a form of the key. However, in general case of mixed quantum states, $V(S_\rho)$ is strictly larger than zero.

We then represent each link in the network by the same state $\rho$ and study its usefulness in the context of hacking. The efficiency of a given scheme we quantify by the difference between the key that can be repeated $R$ and the initial key of the link $K_D$. We then say that a scheme is $(\theta, \eta)$-*good*, when $K_D \geq \eta$ but $R \leq \theta$, along with assumption $\eta > \theta$. This means that the link provides security and because it is not realized by pure state, one can not abuse the link to connect with someone else in the network.

We prove the general lower bound showing that for any state serving as reasonable secure network scheme at least half of the memory qubits (approximately) shall not be used for key distillation, i.e., $V(S_\rho) \geq \frac{1}{2}M(\rho)$. Different, however asymptotically equivalent bound we obtain for the private states [18,22]. For these specific states, we prove that the shield must be at least the size of the key part to assure the security of the scheme. We do so by finding explicit formula for the coherent information of a private state [24,25].

Aiming at set of states for which there are known examples that assure an $\approx (0, 1)$-good scheme, we consider states that have positive partial transposition (PPT states), and approximate some private states. More precisely, we provide lower bounds for the memory cost of our secure network schemes (hubs) employing PPT states approximating strictly irreducible private states [26]. As a related problem being of independent interest, we provide an upper bound on two-way repeater rate for PPT states. These states (i) approximate strictly irreducible pdits for any dimension of the key part $d_k$ (ii) satisfy structural constraints on its behavior under partial transpostion map. For the considered class of states, the overhead approaches 1 in the limit of large dimensions.

However, the speed of this convergence is rather modest. We conclude from the formulas, that e.g., for a scheme with 80% gap, i.e., where $\theta - \eta \geq \frac{8}{10}$, it suffices to spend eight qubits on shield for one qubit in the key part. States realizing such schemes are known [22].

As a byproduct, we prove a lower bound for the trace norm distance between private states and PPT states approximating them. So far, only $d_k = 2$ case was known, which we also tighten. Finally, let us stress that, to our knowledge, the hybrid architecture of a quantum network proposed here is the first application of states with low distillable entanglement (or even bound entangled [16]) in practice.

The paper is organized as follows. In the next section II, we specify and describe an example of the proposed secure-network scheme. In Sec. III, we introduce the memory overhead of the scheme and the density of key. In subsequent Sec. IV, we provide lower bound on overhead for irreducible privates states and also a general lower. In Sec. V, we quantify the scheme that uses private states hardly distinguishable from their attacked versions, whereas in Sec. VI, we concentrate on bounds for certain PPT states. Section VII is left for discussion.

## II. STAR-SHAPED NETWORK: THE CASE STUDY OF THE ATTACK AND COUNTERMEASURE

In this section, we describe in detail the scenario for which, given quantum Internet happens to be realized in a form suggested nowadays, an attack via malware could be done. We then describe countermeasure invoking recent results on limitations on quantum key repeaters

### A. Attack on the star-shaped, pure entanglement based quantum network.

We focus the following specific example of the above-explained scenario. The hub shares secure links with many end-users $E_i$ with $i \in \{1, \ldots, n\}$, in particular with Adam and Eve [see Fig. 2(a)]. The natural topology of the network of secure links is the *star* one (see Fig. 1), so that each end-user is connected with the hub. The hub network node is assumed to be a unit with classical and quantum computer inside. The crucial observation is that if the links are quantum, and based on pure entanglement, they allow via entanglement swapping for the change of topology of the network. Indeed, it can change from star to a disconnected graph of at least two components: star without some nodes and a pair of end-users having a secure connection between them and sharing no more the connection with the hub.

For the above reason, setting up a star network based on pure entanglement, the hub opens a possibility to provide security to pairs of end-users [see Fig. 2(b)]. On the other hand, states allowing quantum communication seem to be an overkill in the case where the node exchanges with subnodes inherently classical information like in the mentioned list of examples of online services. Such an additional side-effect possibility should be under control of the hub that owns the subnetwork. A solution would be to designate a person who sells the connections. If it is not the case, there is a possibility of two dangers: firstly, the administrators of the hub can sell the secure links by themselves and earn illegally without notice of the hub's owner. Secondly, a more dangerous threat is possible: two end-users Adam and Eve can hack the system installing a Trojan quantum software, which serves them as a source of cheap security. Even more importantly, in this way energy consumed for performing quantum operations would be stolen, again, without notice of the hub. Let us note that the same holds if the hub is one of a number of repeater stations [4], and the links are improved via entanglement distillation [9].

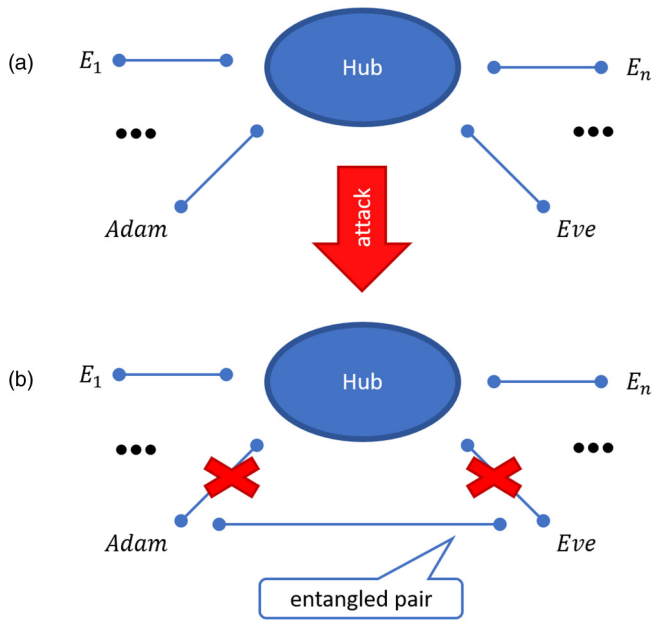

FIG. 2. The main idea of an attack: (a) the hub shares entangled pairs with end-users $E_i$, in particular, with Adam and Eve. Adam and Eve can attack the hub via quantum malware which performs for them entanglement swapping (b) Adam and Eve share an entangled pair after successful attack.

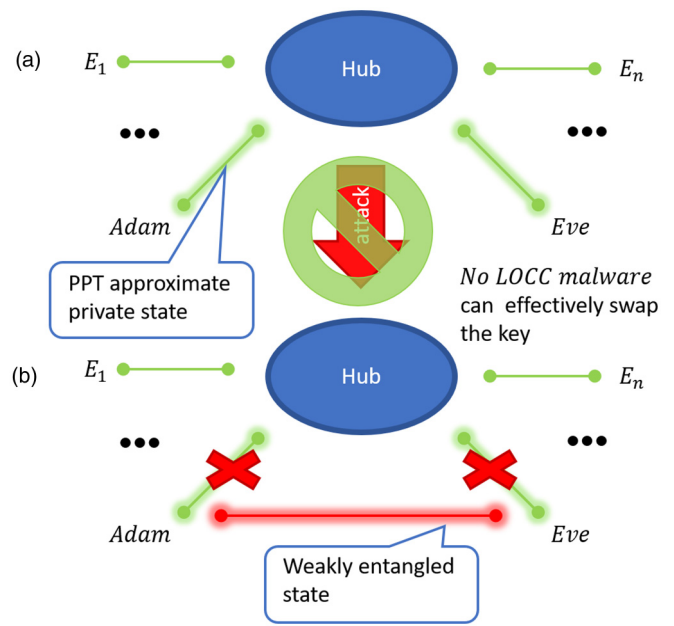

FIG. 3. The main idea of the countermeasure: (a) hub shares bound entangled states (green lines) with end-users $E_i$ (each having at least 1 bit of key), in particular, with Adam and Eve (shaded lines). No LOCC malware can efficiently swap the key. (b) Adam and Eve can not share a state (shaded red line) with non-negligible amount of key (compare [15]).

We will distinguish here two kinds of attacks: a *general one* where the hacked node can perform any three-party classical communication with two other nodes and *one-way attack* where only the central node can communicate classical information to the hackers.

### B. Countermeasure via noisy entangled states

In what follows, we observe, that using appropriate noisy entangled states solves the mentioned problem in cases of general (two-way) and one-way attack (see Fig. 3).

Recently a fundamental result has been shown in this context, indicating that for some states (having at least one separable key attacked state) the rate $R$ of repeated secure key is strongly related to the so-called *distillable entanglement* [9,17,19] by the following result:

$$R^{H_1H_2 \to A:E}(\gamma_{AH_1}, \gamma_{H_2E}) \leqslant E_D^{H_1H_2 \to A:E}(\gamma_{AH_1} \otimes \gamma_{H_2E}), \quad (2)$$

where $\to$ stands for the classical communication restricted to one-way from the intermediate node $H \equiv H_1H_2$ to nodes $A$ and $E$, and $\gamma_{AH_i}$ denotes a private state [18]—a state possessing ideal security directly accessible via measuring its subsystem called *key part*.

*Notation 1.* Private state with $d_k$ dimensional key part and $d_s$ dimensional shield part per one party, shared between $A$ or $E$ and $H$ is denoted $\gamma_{d_k,d_s}$.

We present the following countermeasure: instead of having a star network with the end-users, which is pure entanglement based, the hub can set a star-shaped network of point-to-point links based on bound entangled states which are approximate private states (see Fig. 3). Let us note that this is legitimate when the hub needs to encrypt only classical data. Furthermore, if end-users had a quantum connection with the hub, then we could have a case of hub's network abuse. These bound entangled states are *weakly* transitive. This means that there *does not exist* a quantum software that can be run on the quantum computer of the hub, or even a quantum tripartite LOCC protocol between Adam, Eve, and the quantum computer of the hub, to achieve this task. The *no-go* is hence turned into a success. The hub employing the bound entanglement based quantum links keeps secure communication but needs not to control the setup. There is *no* physical map in the considered scenario, that can create a secure link with a non-negligible amount of secrecy.

Let us note that although we have mentioned here entanglement swapping, in Ref. [15], it is shown that even if the links with the hub are provided in the form of a private state $\gamma_{AH_i}^{\otimes n}$ (or an approximate private state), the rate of the output secure key for Adam and Eve is negligible as a function of dimension of the bound entangled approximate private states. By *negligible* amount, we mean the rate which goes to zero with growing dimension of the shield system of the state $\gamma_{AH_i}$. Hence the countermeasure works in the asymptotic regime up to the fact that some small rate of key can be obtained by Adam and Eve. Note here that the key of the links is not in the form of pure entanglement. The states are chosen such that they have negligible or even zero distillable entanglement. However, such a choice of states for the links is not a constraint for the security of the network. This is because we consider only hubs that send classical data, hence their encryption does not need to involve pure entanglement. Moreover, the states considered in our solution, have large enough distillable key.

## III. MEMORY OVERHEAD OF THE COUNTERMEASURE

We now focus on the *quantum memory cost* of implementation of the proposed countermeasure. We recall first the definition of the key repeater rate. Let us stress here that according to our approach, the lower it is, the better for the security of the node.

We further focus on the scheme represented by a private state with $d_k$ dimensional key-part and $d_s$ dimensional shield. This state reads a form [18]:

*Definition 1.* Private quantum state

$$\gamma_{d_k,d_s} := \sum_{i,j=0}^{d_k-1} \frac{1}{d_k} |ii\rangle\langle jj| \otimes X_{ij}, \quad (3)$$

where $X_{ij} = U_i \sigma U_j^\dagger$ for some state $\sigma$ of $\mathcal{C}^{d_s} \otimes \mathcal{C}^{d_s}$ and $U_i$ are some unitary transformations.

*Notation 2.* We follow the notation in which

$$\|X\|_1 = \mathrm{Tr}\sqrt{XX^\dagger}. \quad (4)$$

Additionally we skip the subscript, as it doesn't lead to any ambiguity.

*Remark 1.* Through the rest of the paper, we assume that each considered quantum state $\rho$ acts on $\mathcal{H}_H \otimes \mathcal{H}_N$ being tensor product of subspaces associated with the hub and a node, and $\dim \mathcal{H}_H = \dim \mathcal{H}_N < \infty$. What is more, both subspaces are assumed to be partitioned into key and shield parts (of dimensions $d_k$ and $d_s$, respectively) in the same way at both sides, unless stated differently.

*Notation 3.* Here we adapt shortened notation in which $X_{ij} \equiv X_{ii,jj}$. In calculations, we mainly incorporate full notation. Additionally for $i \neq j$, we define $X_{ij,ij} \equiv 0$, as they do not enter to definition of a private state.

Note that $X_{ii}$ are, in fact, subnormalized states, obtained on the shield system upon observing key $|i\rangle$ on the key part. We call them *conditional states*. According to definition, $K_D(\gamma_{d_k,d_s}) \geqslant \log_2 d_k$, while in case of equality, a private state is called *irreducible*: its whole secure content is available from the key part via direct measurement. In the case in which $X_{ii}$ are additionally separable, we call these states strictly irreducible private states. In fact, it is conjectured that all irreducible private states are of the form of strictly irreducible ones [26], it is so if there do not exist entangled but key-undistillable states.

*Definition 2.* The distillable key rate with respect to arbitrary LOCC operations is defined as

$$K_D(\rho) := \inf_{\epsilon > 0} \limsup_{n \to \infty}$$

$$\sup_{\Lambda_n^{\mathrm{LOCC}}, \gamma_{d_k,d_s}} \left\{ \frac{\log_2 d_k}{n} : \Lambda_n^{\mathrm{LOCC}}(\rho^{\otimes n}) \approx_\epsilon \gamma_{d_k,d_s} \right\}, \quad (5)$$

where $\rho$ is a bipartite state shared by the parties. $\Lambda$ is a LOCC protocol with two-way classical communication.

*Definition 3.* The quantum key repeater rate with respect to arbitrary LOCC operations among $A$, $E$, and $H$ is defined as

$$R^{A \leftrightarrow H \leftrightarrow E}(\rho, \rho') := \lim_{\epsilon \to 0} \lim_{n \to \infty}$$

$$\sup_{\Lambda_n^{\mathrm{LOCC}}, \gamma_{d_k,d_s}} \left\{ \frac{\log_2 d_k}{n} : \mathrm{Tr}_H \Lambda_n^{\mathrm{LOCC}}((\rho \otimes \rho')^{\otimes n}) \approx_\epsilon \gamma_{d_k,d_s} \right\}, \quad (6)$$

where Adam and Hub share state $\rho$ while Hub and Eve share $\rho'$. $\Lambda := \{\Lambda_n^{\mathrm{LOCC}}\}$ are tripartite LOCC protocols with two-way classical communication. In the case in which communication between central node and $A$, $E$ systems is restricted to one-way from $H$ to $A$ and $E$, we denote this rate with $R^{H \to A:E}$.

*Notation 4.* For the repeater rate in the case in which $\rho = \rho'$, we introduce simplified notation $R^{\to(\leftrightarrow)}(\rho)$.

An ultimate goal would be to provide a nonrepeatable key with the smallest possible memory cost, being a precious resource in NISQ era of quantum computing. Our solution to the problem is represented by a bipartite quantum state $\rho$ shared between the central node $H$ and one of the end-users (Adam), however, its specific parameters are important enough to write them out explicitly. The scheme will be represented by the following tuple:

$$S_\rho^{\to(\leftrightarrow)} := \langle \rho, \log_2 \dim_H(\rho), \Delta, K_D(\rho), R^{\to(\leftrightarrow)}(\rho) \rangle. \quad (7)$$

The arrow(s) in the superscript are dropped if the results hold for both cases. The state $\rho_{HA}$ is shared between the central node $H$ and a single end-user (Adam). $\Delta$ is the degree of the node (number of connections).

*Definition 4.* The scheme $S_\rho$ is one-way (two-way) $(\theta, \eta)$-good if $R^{\to(\leftrightarrow)}(\rho) \leqslant \theta$ and $K_D(\rho) \geqslant \eta$, and $\eta > \theta$.

Let us note here, that by definition of the key repeater rate, $K_D \geqslant R^\leftrightarrow \geqslant R^\to$. This means that if $R^\leftrightarrow = \theta$, we have $\theta \leqslant \eta$. However, since $\theta$ is only an upper bound on $R^\leftrightarrow$, we had to assume in the above definition desired order of parameters. Moreover we demand strict inequality $\eta > \theta$, since the scheme with $\eta = \theta$ is not "good," i.e., does not have any advantage over quantum repeaters design. Indeed, in our approach, we are interested in the largest possible gap between $K_D$ and $R^{\to(\leftrightarrow)}$, while keeping memory overhead considerably small at the same time. We quantify this gap by its lower bound defined as a difference $\eta - \theta$, and call it *efficiency* of the scheme.

*Definition 5.* The *overhead* of the scheme $S_\rho$ is the following quantity:

$$V(S_\rho) := \Delta(\log_2 \dim_H(\rho) - K_D(\rho)), \quad (8)$$

where $\rho$ is a bipartite state shared between the hub and a end-user.

The overhead is the difference between the qubits of memory at the node: $\rho^{\otimes \Delta}$ has

$$M(\rho) := \Delta \log_2 \dim_H(\rho), \quad (9)$$

of qubits of subsystem $H$, and the number of bits of security which the node shares with the other part of the quantum Internet.

*Definition 6.* For a scheme that is $(\theta, \eta)$-good, the difference $\eta - \theta \geqslant 0$ is the gap of the scheme.

We note here that such defined overhead bares strong connection with the other, to our knowledge not explicitly studied notion, which is that of *density of the private key*.

*Definition 7.* For a quantum state $\rho$ ($\dim_H(\rho) \geqslant 2$) shared between the hub $H$ and Adam $A$ or Eve $E$, the density of the private key $\mathcal{D}$ reads

$$\mathcal{D}(\rho) := \frac{K_D(\rho)}{\log_2 \dim_H(\rho)}. \quad (10)$$

In the above definition, we have included a restriction $\dim_H(\rho) \geqslant 2$ to exclude trivial, not relevant cases. We then have the dependence

$$V(S_\rho) = M(\rho)(1 - \mathcal{D}(\rho)). \quad (11)$$

From the above form it is clear to see that the overhead is a non-negative quantity, as the density is a quantity less than or equal to 1. In what follows, we provide several lower bounds on the overhead $V$ of the countermeasure, that satisfies

$$0 \leqslant V(S_\rho) \leqslant M(\rho). \quad (12)$$

The first from the above inequalities follows from the fact that secure key $K_D(\rho)$, can not be larger than memory size $\log_2 \dim_H(\rho)$, and hence $\Delta$ is non-negative. Presenting results on plots in the next sections, we will concentrate on the fraction between memory overhead and total memory of the hub, i.e., the percentage of total memory that is not used for storing secret-key.

## IV. LOWER BOUNDS ON THE OVERHEAD OF THE SECURE NETWORK SCHEME

Let us first focus on the class of *one-way attacks*: the attacked hub node can send data to two receiver nodes owned by malicious parties that can communicate freely. We begin with preliminary definitions and facts.

*Definition 8.* The coherent information of a quantum state $\rho_{AB}$

$$I_{\text{coh}}(A \rangle B) = S(B) - S(AB), \quad (13)$$

where $S(B)$ is the Von Neumann entropy of state $\rho_B = \text{Tr}_A(\rho_{AB})$ and $S(AB)$ is that of state the $\rho_{AB}$.

The key repeater rate is an upper bound on distillable entanglement in each of the two links of the star-shaped network. We therefore provide a lower bound on one-way distillable entanglement $E_D^{\rightarrow}(.)$ of a private state, via the Devetak-Winter hashing protocol [27].

$$E_D^{\rightarrow}(\gamma_{d_k,d_s}) \geqslant \log_2 d_k + \sum_i \frac{1}{d_k} I_{\text{coh}}(A' \rangle B')_{\sigma_i}, \quad (14)$$

where $I_{\text{coh}}$ is the coherent information [17], and $\sigma_i$ are the conditional states of a private state. We have then the following observation.

*Observation 1.* For any private state $\gamma_{d_k,d_s}$, one-way distillable entanglement is lower bounded as follows:

$$E_D^{\rightarrow}(\gamma_{d_k,d_s}) \geqslant \log_2 d_k + \sum_i \frac{1}{d_k} I_{\text{coh}}(A' \rangle B')_{\sigma_i}, \quad (15)$$

where $\sigma_i = U_i \rho_{A'B'} U_i^\dagger$ are conditional states.

For the proof of the above observation see Appendix.

Let us note that the above bound is achievable given a choice $\forall_i \sigma_i = \frac{I}{d_s}$, i.e., for pdits with twisted-in maximally mixed state.

Since coherent information can not be smaller than $-\log_2 d$ for a $d$ dimensional state, we have the following general result.

*Corollary 1.* For any private state $\gamma_{d_k,d_s}$ one-way distillable entanglement is lower bounded by the following expression:

$$E_D^{\rightarrow}(\gamma_{d_k,d_s}) \geqslant \log_2 d_k - \log_2 d_s, \quad (16)$$

where $d_k$ and $d_s$, are dimensions of the key part and shield part respectively.

*Proof.* It follows from the fact that for any state $\sigma_i$ of dimension $d_s^2$, there is $I_{\text{coh}}(A' \rangle B')_{\sigma_i} \geqslant -\log_2 d_s$. Indeed, $S(B') - S(A'B') = I(A' : B') - S(A') \geqslant 0 - \log_2 |A'| = -\log_2 d_s$, as the entropy is maximally $\log_2 |A'|$ while $I(A' : B') \geqslant 0$. ∎

Following the fact that one-way distillable entanglement constitutes a lower bound on both one-way and two-way repeater rates, we conclude that in schemes incorporating privates states, it is reasonable to assume $d_s \geqslant d_k$. This assumption is a necessary condition for having low repeater rates.

As we have discussed, we obtain the following lower bound on overhead of schemes based on irreducible private states.

*Theorem 1.* If an irreducible private state $\gamma_{d_k,d_s}$ serves as an $(\theta, \log_2 d_k)$-good secure network scheme $S^{\rightarrow}_{\gamma_{d_k,d_s}}$ with degree $\Delta$, then its overhead satisfies a lower bound:

$$V(S^{\rightarrow}_{\gamma_{d_k,d_s}}) \geqslant \Delta \log_2(d_k d_s)\left(1 - \frac{1}{2 - \frac{\theta}{\log_2 d_k}}\right) \quad (17)$$

$$\approx_{\theta \approx 0} \frac{1}{2} M(\gamma_{d_k,d_s}). \quad (18)$$

For the proof of the above theorem see Appendix.

This theorem shows that memory used by a private state which allows only for $\theta$ of repeated key must have at least as big shield system as its key part, see Fig. 4 for exemplary lower bounds. The technique used for proving theorem 1 inspired us to find a general lower bound on the overhead of any scheme, which is presented below.

*Theorem 2.* Any state $\rho$ that serves as $(\theta, \eta)$-good secure network scheme, satisfies

$$V(S_\rho) \geqslant M(\rho)\left(\frac{1}{2} - \frac{\theta}{\log_2 d_H}\right) \approx_{\theta \approx 0} \frac{1}{2} M(\rho). \quad (19)$$

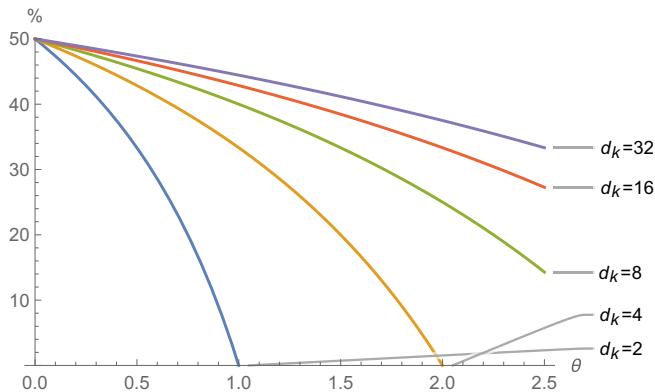

FIG. 4. Plots of lower bounds on percentage of memory overhead from theorem 1, with different values of $d_k$.

For the proof of the above theorem see Appendix. The above theorem is based on observation that distillable key is upper bounded by $S(A)/2$ if only coherent information is nonpositive. As we will show below on Fig. 5, this bound is the only bound on key repeater rate for certain amount of one-way distillable entanglement.

The bound shown in Fig. 5 as dotted blue line segment reads

$$R^{\rightarrow}(\rho) \leqslant R^{\leftrightarrow}(\rho) \leqslant K_D(\rho)$$
$$\leqslant E_{\mathrm{sq}}(\rho) \leqslant \frac{S(A)}{2} + \frac{E_D^{\rightarrow}(\rho)}{2}. \quad (20)$$

The inequality in Eq. (20) is a known fact, since one-way communication from the hub $H$ to hosts $A$ and $E$ can not allow to repeat more key than in two-way communication setup. The second inequality comes from the fact that it is not possible to have more of a repeatable key than a distillable key. On the other hand, it is possible that the quantum key repeater rate is smaller than the distillable key of a particular state $\rho$. The third inequality is true because squashed entanglement is an upper bound on distillable key [28]. Finally, the last inequality is the upper bound on $R^{\rightarrow}$ observed in this work, which is a direct consequence from the proof of lemma 18 in Ref. [15]. Similar results on private capacity for quantum channels were obtained in Ref. [29]. As one can see the result for states which we prove here is much simpler than the analogous one proved there for channels.

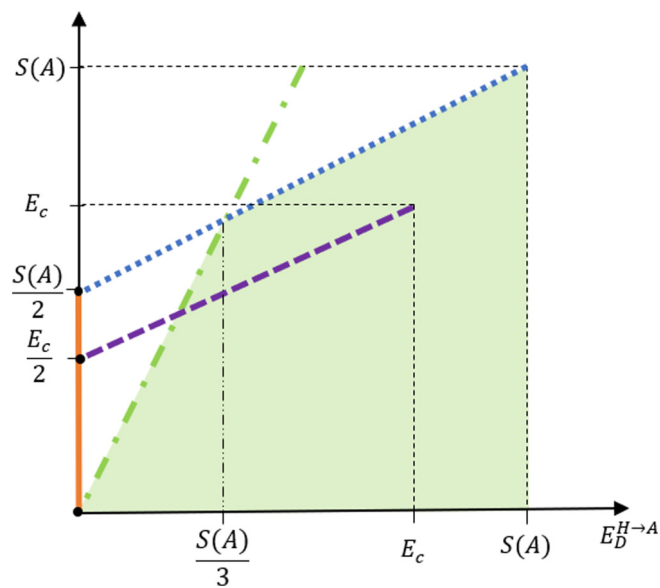

FIG. 5. Upper bounds on quantum key repeater rates. Dotted blue line: introduced here in eqn. (20), solid orange line: special case for $I_{\mathrm{coh}} = 0$, dashed violet line: [15], and dotted-dashed green line: only for some states [19]. Shaded region corresponds to a combination of bounds.

The dotted-dashed green line segment is the upper bound on quantum key repeater rate derived in Ref. [19]:

$$R^{\rightarrow}(\rho) \leqslant 2E_D^{\rightarrow}(\rho), \quad (21)$$

which holds for special class of *block states*. Here the hub can send messages to Adam and Eve, but not receive from them. Adam and Eve can communicate in both ways freely.

The dashed violet line segment is the upper bound introduced in Ref. [15]:

$$R^{A \leftarrow H \rightarrow E}(\rho) \leqslant \frac{E_D^{\rightarrow}(\rho)}{2} + \frac{E_C(\rho)}{2}. \quad (22)$$

In this case, only the communication from Hub to Adam is one-way and between Hub and Eve the communication is one-way, no other data transfer is allowed.

The solid orange line segment is the upper bound for states that have $I_{\mathrm{coh}} = 0$. These states do not have more of distillable key than $E_C/2$ or $S(A)/2$.

Even though dotted-dashed green and dashed violet bounds intersect in $E_D^{\rightarrow} = E_c/3$, they are different scenarios in which the classical communication is not in the same direction. Therefore, they are incomparable. It is the same for dotted blue and dashed violet bounds. On the other hand, the directions of classical communication for dotted-dashed green and dotted blue bounds are the same, so it is possible to compare them. The upper bound introduced in this work is more accurate than the bound derived in Ref. [19] starting from $E_D^{\rightarrow} = S(A)/3$.

## V. LOWER BOUND ON OVERHEAD FOR PRIVATE STATES HARDLY DISTINGUISHABLE FROM THEIR ATTACKED VERSIONS

In this section, we derive lower bounds for the memory overhead for schemes utilizing private states hardly distinguishable from their attacked versions. We first briefly explain the approach and then formalize the presented idea.

Let us note, that to assure $\eta > 0$ in our scheme $S_{\rho}$, we need to know how much a given state of it $\rho$ has distillable key. A good choice is then a strictly irreducible private state, as for this state, we know that $K_D(\gamma_{\langle d_k, d_s \rangle}) = \log_2 d_k$, however, such $\gamma_{\langle d_k, d_s \rangle}$ should not be too much distillable, as $R^{\leftrightarrow} \geqslant E_D(\rho)$. Thus, to also have that scheme is $(\theta, \eta)$-good for small $\theta$, we need to assure $E_D(\gamma_{\langle d_k, d_s \rangle}) \leqslant \theta$. This can be done in various ways, including logarithmic negativity bound $E_D(\rho) \leqslant -\log_2 \|\rho^{\Gamma}\|$ [30]. From Ref. [19], it follows that $R^{\rightarrow}$ is small since it is upper bounded by $2E_D^{\rightarrow}$. The next theorem encapsulates this approach and proves the lower bound on the memory cost of such a solution.

We first use the bound that employs measure called log-negativity [30,31].

*Observation 2.* For a private state such that $X_{ii} \in PPT$, and at least one from its conditional key attacked states is separable, there is the following bound on the one-way quantum key repeater rate:

$$E_D^{\rightarrow}(\gamma_{d_k,d_s} \otimes \gamma_{d_k,d_s}) \leqslant 2\log_2\left(1 + \|\gamma_{d_k,d_s}^{\Gamma} - \hat{\gamma}_{d_s,d_k}^{\Gamma}\|\right), \quad (23)$$

where $\hat{\gamma}_{d_k,d_s} = \sum_i \frac{1}{d_k}|ii\rangle\langle ii| \otimes X_{ii}$ is an irreducible private state after measurement on the key part (attacked), and $\Gamma$ is an operation of partial transposition.

For the proof of the above observation see Appendix.

For technical reasons, we deal more specifically with the right-hand side of the above inequality, as encapsulated in the following observation.

*Observation 3.* The following identity holds:

$$\|\gamma_{d_k,d_s}^{\Gamma} - \hat{\gamma}_{d_k,d_s}^{\Gamma}\| = \sum_{i \neq j} \frac{1}{d_k}\|X_{ij}^{\Gamma}\|, \quad (24)$$

where $\hat{\gamma}_{d_k,d_s}^{\Gamma} = \sum_i \frac{1}{d_k}|ii\rangle\langle ii| \otimes X_{ii}^{\Gamma}$ is the private state after measurement on the key part and $\Gamma$ is the partial transpose operation.

For the proof of the above observation see Appendix.

In the next lemma, we argue that some private states, that are hardly distinguishable from their attacked versions, have large dimension of the shield in relation to the dimension of the key part.

*Lemma 1.* For a special private state $\gamma_{d_k,d_s}$, which satisfies condition $X_{ii}^{\Gamma} \geqslant 0$, and $\|\gamma_{d_k,d_s}^{\Gamma} - \hat{\gamma}_{d_k,d_s}^{\Gamma}\| \leqslant \epsilon$, there is

$$d_s \geqslant \frac{d_k - 1}{\epsilon}. \quad (25)$$

For the proof of the above lemma see Appendix.

The above technical lemma and observation lead us to the main result of this section. It states that the overhead in case of private states that are hardly distinguishable from their attacked versions tends to 1 with the parameter of distinguishability approaching zero.

*Theorem 3.* A strictly irreducible private state $\gamma_{\langle d_k,d_s \rangle}$ ($X_{ii} \in SEP$, $d_k \geqslant 2$) satisfying $\|\gamma^{\Gamma}_{\langle d_k,d_s \rangle} - \hat{\gamma}^{\Gamma}_{\langle d_k,d_s \rangle}\| \leqslant \epsilon$ and $\frac{d_k-1}{d_s} \leqslant \epsilon$ serves as $(\theta, \eta)$-good secure network scheme scheme with

$$V(S_{\gamma_{\langle d_k,d_s \rangle}}^{\rightarrow}) \geqslant M(\gamma_{\langle d_k,d_s \rangle})\left(1 - \frac{\log_2 d_k}{\log_2 d_k + \log_2 \frac{d_k-1}{\epsilon}}\right) \quad (26)$$

$$\approx_{\epsilon \to 0} M(\gamma_{\langle d_k,d_s \rangle}), \quad (27)$$

for $\theta = 2\log_2(1+\epsilon) \approx_{\epsilon \ll 1} \frac{2}{\ln 2}\epsilon$ and $\eta = \log_2 d_k$.

For the proof of the above theorem see Appendix. For the performance of lower bound in different dimensions of key part see Fig. 6, for the behaviour of a gap see Fig. 7.

### A. Example of the gap for low dimensional state

In general, one would like to diminish the repeater rate of the scheme as much as possible. Unfortunately, in theorem 3,

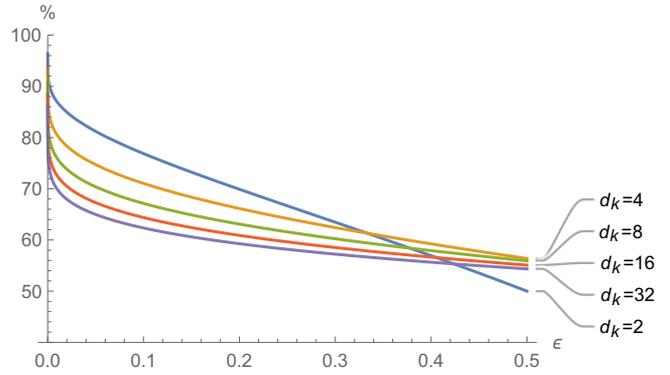

FIG. 6. Plots of lower bounds on percentage of memory overhead from theorem 3, with different values of $d_k$.

the parameter $\epsilon$ appears both in formula for the repeater rate and the overhead. This is the reason why one can not reduce repeater rate to zero keeping the overhead smaller than total memory cost. In this situation, one should decide on an acceptable level of repeater rate, for which the overhead is still reasonable. A small dimensional example of a pbit state which allows for such a control is known [18,32]. Block matrix representation of such a pbit is

$$\Omega_{d_s} = \frac{1}{2}\begin{bmatrix} \frac{I}{d_s^2} & 0 & 0 & \frac{F}{d_s^2} \\ 0 & 0 & 0 & 0 \\ 0 & 0 & 0 & 0 \\ \frac{F}{d_s^2} & 0 & 0 & \frac{I}{d_s^2} \end{bmatrix}, \quad (28)$$

where F is a matrix of swap quantum logic gate of dimension $d_s^2$ implying $\|\Omega_{d_s}^{\Gamma} - \hat{\Omega}_{d_s}^{\Gamma}\| = \frac{1}{d_s}$. We estimate now the size of the gap for a scheme using this state. Let us assume a scheme with minimal amount of memory by setting $\epsilon = \frac{1}{d_s}$ (see that conditions of lemma 1 and theorem 3 are satisfied). We obtain a lower bound $V(S_{\gamma_{\langle d_k,d_s \rangle}}) \geqslant M(\gamma_{\langle d_k,d_s \rangle})(1 - \frac{1}{1+\log_2 d_s})$, for scheme being $(\frac{2}{\ln 2}\frac{1}{d_s}, 1)$-good. For $d_s = 2$, it saturates also the general lower bound on overhead from theorem 2 with value of $\frac{1}{2}$, although in this case the rate of repeater $R^{\rightarrow}$ is upper bounded with $\frac{1}{\ln 2} \approx 1.44$, what is an unsatisfying result. The first nontrivial case, in that secure network scheme

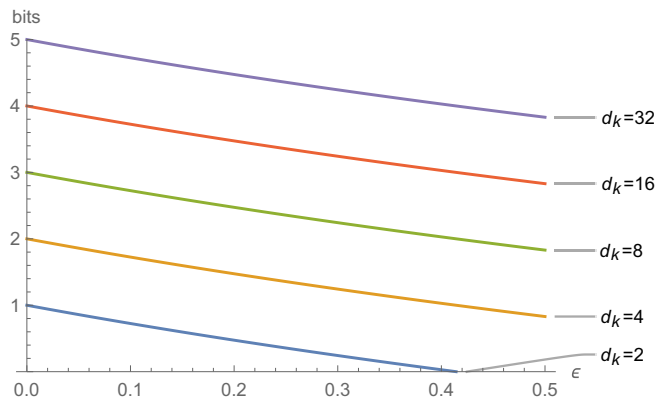

FIG. 7. Plots of lower bound on gap between $\eta$ and $\theta$ for the scheme in theorem 3.

has an advantage over malicious parties, appears for $d_s = 3$, in which repeatable rate drops to $R^{\rightarrow} \leqslant \frac{2}{3 \ln 2} \approx 0.96$ being strictly smaller than key rate $K = 1$, what follows from its irreducibly.

## VI. LOWER BOUNDS ON OVERHEAD FOR PPT STATES

As it was argued in Ref. [15] (see supplemental material note 6) the states which are PPT and approximate private bits are of rather high dimension. This fact can be found as a consequence of the following earlier statement [23] (see also Ref. [33]):

$$\forall_{\rho \in PPT \gamma \in \mathcal{C}^2 \otimes \mathcal{C}^2 \otimes \mathcal{C}^{d_s} \otimes \mathcal{C}^{d_s}} \|\rho - \gamma_{2,d_s}\| \geqslant \frac{1}{2(d_s+1)}. \qquad (29)$$

We conclude that a quantum PPT state close by $\epsilon$ in the trace norm to strictly irreducible private state $\gamma_{d_k,d_s}$ has dimension of the shield at least $d_s \geqslant \frac{1-2\epsilon}{2\epsilon}$.

It is known that for two-way repeater rate to be zero, the state has to be bound entangled [$R^{\leftrightarrow}(\rho) \geqslant E_D(\rho)$] [18]. Thus, in this section, we investigate the overhead using such schemes.

*Notation 5.* We adopt a notation in which PPT state $\rho$ has the following form:

$$\rho := \sum_{i,j,k,l=0}^{d_k-1} |ij\rangle\langle kl| \otimes A_{ij,kl}, \qquad (30)$$

where come $A_{ij,kl}$ are blocks of dimension $d_s^2$.

*Proposition 1.* If $\rho$ is a state with positive partial transpose, that approximates a strictly irreducible private bit $\|\rho - \gamma_{\langle 2,d_s \rangle}\| \leqslant \epsilon$ for $\epsilon \geqslant \frac{1}{2(d_s+1)}$, $\|A^{\Gamma}_{01,10}\| \leqslant \epsilon$, and its conditional shield states are separable, then its two-way repeater rate $R^{\leftrightarrow}(\rho)$ is upper bounded as follows:

$$R^{\leftrightarrow}(\rho) \leqslant 2\left(\sqrt{\epsilon} + \frac{3}{2}\epsilon\right)(1 + \log_2 d_s) \\ + (1 + 2\sqrt{\epsilon} + 3\epsilon)h\left(\frac{2\sqrt{\epsilon} + 3\epsilon}{1 + 2\sqrt{\epsilon} + 3\epsilon}\right). \qquad (31)$$

For the proof of the above proposition see Appendix.

Note that PPT states from proposition 1 above do exist. One example can be states for which $A_{01,10} = A^{\Gamma}_{00,11}$. For upper bounds on key repeater rate from this proposition for dimensions $d_s = 2, \dots, d_s = 32$ see Fig. 8.

*Theorem 4.* If a state with positive partial transpose $\rho$ approximates strictly irreducible private bit $\|\rho - \gamma_{\langle 2,d_s \rangle}\| \leqslant \epsilon$ for $\frac{1}{2(d_s+1)} \leqslant \epsilon < \frac{1}{2}$, $\|A^{\Gamma}_{01,10}\| \leqslant \epsilon$, and its conditional shield states are separable, then it serves as a two-way $(\theta, \eta)$-good secure network scheme $S_\rho$ with degree $\Delta$, and its overhead satisfies a lower bound:

$$V(S_\rho) \geqslant M(\rho)\left(1 - \frac{1 + (1+\frac{\epsilon}{2})h\left(\frac{\frac{\epsilon}{2}}{1+\frac{\epsilon}{2}}\right)}{1 + \log_2\left(\frac{1-2\epsilon}{2\epsilon}\right)} - \frac{\epsilon}{2}\right) \qquad (32)$$

with $\eta = 1 - 8\epsilon - 4h(\epsilon)$ [where $h(.)$ is the binary Shannon entropy] and $\theta = 2(\sqrt{\epsilon} + \frac{3}{2}\epsilon)(1 + \log_2 d_s) + (1 + 2\sqrt{\epsilon} + 3\epsilon)h(\frac{2\sqrt{\epsilon}+3\epsilon}{1+2\sqrt{\epsilon}+3\epsilon})$.

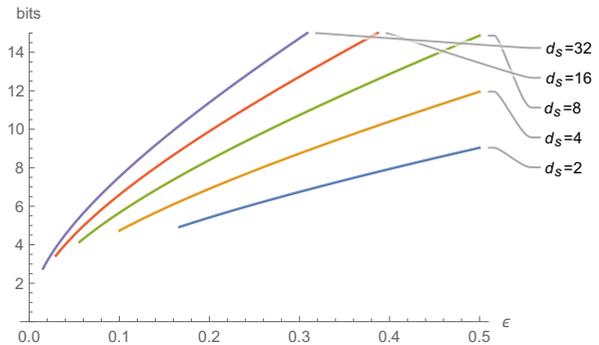

FIG. 8. Upper bounds on key repeater rate from proposition 1. Domains are constrained with $\epsilon \geqslant \frac{1}{2(d_s+1)}$ condition.

For the proof of the above theorem see Appendix. For the plot of lower bound of percentage of memory overhead from the above theorem see Fig. 9. The plots of lower bounds on gap between $\eta$ and $\theta$ in this theorem are depicted in Fig. 10. From inequality (29), we obtain

$$\log_2 d_s \geqslant \log_2\left(\frac{1-2\epsilon}{2\epsilon}\right). \qquad (33)$$

We then see that focusing on states which have positive partial transposition and approximate private bits is quite costly: the overhead approximates the whole memory of the scheme for small $\epsilon$. In particular, obtaining a reasonable amount of key in links $\approx 1$ bits for each of $\Delta$ links implies that the whole memory cost is that of an overhead. However, an advantage of this scheme is that it is no longer limited to one-way communication. In this case, there *does not exist any three-partite LOCC protocol which can break the scheme*.

We now generalize the above result for larger dimensions of the key part than qubit, and study it in case of private state. In order to achieve this we need a number of technical observations and lemmas, which we present below.

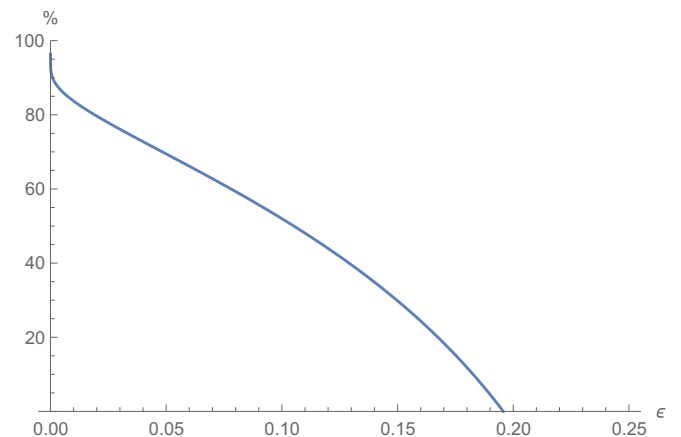

FIG. 9. Plot of lower bound on percentage of memory overhead from theorem 4.

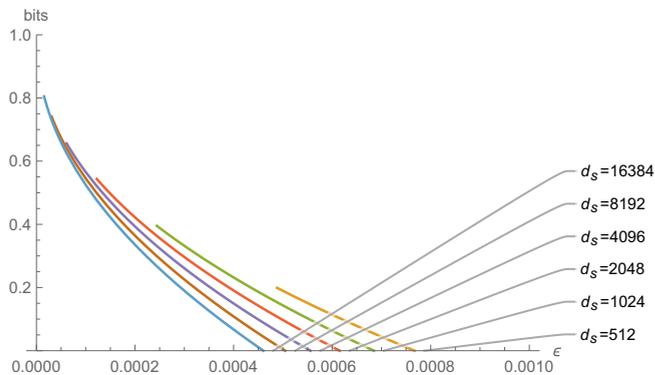

FIG. 10. Plots of lower bounds on gap between $\eta$ and $\theta$ for the scheme in theorem 4. The case $d_s = 64$ is a setting with lowest dimension for obtaining positive lower bound on gap under $\epsilon \geqslant \frac{1}{2(d_s+1)}$ condition.

*Observation 4.* Denoting with $A_{ij,kl}$ matrices some of them ($A_{ii,jj}$) being unnormalized conditional states of the shield of a state $\rho = \sum_{ijkl} |ij\rangle\langle kl| \otimes A_{ij,kl}$, we prove the following relations:

$$\|\rho - \gamma\| \leqslant \epsilon \Rightarrow \forall_{i \neq j} \|A_{ii,jj}\| \geqslant \frac{1}{d_k} - \epsilon \quad (34)$$

and

$$\|\rho - \gamma\| \leqslant \epsilon \Rightarrow \sum_{i \neq j} \|A_{ij,ij}\| \leqslant \epsilon. \quad (35)$$

In the following lemma and subsequent corollary, we prove a general lower bound on the distance between private states (of any dimension of the key part) from PPT states [23].

*Lemma 2.* For any state $\rho \in PPT$, there is

$$\|\rho - \gamma_{d_k,d_s}\| \leqslant \epsilon \Rightarrow d_s \geqslant \left(\frac{d_k - 1}{\epsilon}\right)(1 - \epsilon d_k), \quad (36)$$

where $\gamma$ is a private state with $d_k^2$ dimensional key part and $d_s^2$ dimensional shield subsystem.

*Corollary 2.* For any state $\rho \in PPT$ approximating private state, the following lower bound holds:

$$\|\rho - \gamma_{d_k,d_s}\| \geqslant \frac{d_k - 1}{d_s + d_k(d_k - 1)}. \quad (37)$$

The important properties of lower bound presented in corollary 2 are the fact that it is not trivial for values of $d_k$ but also that it yields tighter bound for $d_k = 2$ known form [23] [see Eq. (29)]. Concluding as a byproduct, we have found a nontrivial (nonzero) lower bound on the distance between any private state and a PPT state in any dimension [23,32].

*Corollary 3.* For any state $\rho \in PPT$ of dimension $2d_s \otimes 2d_s$ approximating private bit there is

$$\|\rho - \gamma_{2,d_s}\| \leqslant \epsilon \Rightarrow \|A^\Gamma_{01,10}\| \leqslant \frac{\epsilon}{2}. \quad (38)$$

The upper bound on the norm in Corollary 3 is tighter than the one in [23]. This is due to modification in the proof technique. This motivates us to assume $\sum_{i \neq j} \|A^\Gamma_{ij,ji}\| \leqslant \epsilon$, instead of $2\sum_{i \neq j} \|A^\Gamma_{ij,ji}\| \leqslant \epsilon$ what would be analogous to assumption in proposition 1.

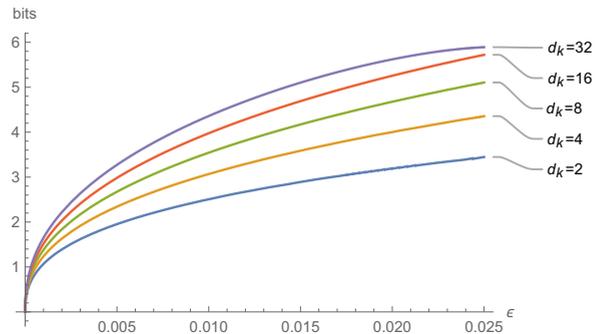

FIG. 11. Upper bounds on key repeater rate from proposition 2. We extract from condition $\frac{d_k-1}{d_s+d_k(d_k-1)} \leqslant \epsilon$ minimal value of $d_s = \lceil \frac{d_k - 1 - \epsilon d_k(d_k-1)}{\epsilon} \rceil$ yielding best value for upper bound.

*Proposition 2.* If $\rho$ is a state with positive partial transpose approximates a strictly irreducible private dit (pdit) $\|\rho - \gamma_{\langle d_k,d_s \rangle}\| \leqslant \epsilon$ for $\frac{d_k-1}{d_s+d_k(d_k-1)} \leqslant \epsilon$, $\sum_{i \neq j} \|A^\Gamma_{ij,ji}\| \leqslant \epsilon$, and conditional shield states of $\rho$ are separable, then its two-way repeater rate $R^\leftrightarrow(\rho)$ is upper bounded as follows:

$$R^\leftrightarrow(\rho) \leqslant 2(\sqrt{\epsilon} + \epsilon) \log_2 \dim_H(\rho)$$
$$+ (1 + 2\sqrt{\epsilon} + 2\epsilon) h\left(\frac{\sqrt{\epsilon} + \epsilon}{\frac{1}{2} + \sqrt{\epsilon} + \epsilon}\right). \quad (39)$$

For the proof of the above proposition see Appendix. It is easy to notice that the upper bound in proposition 2 evaluated for pbits is tighter than the corresponding one from proposition 1. This is because with slightly different assumption on $A_{ij,ji}$ blocks. For upper bounds on the key repeater rate for exemplary dimensions of the key part provided in proposition 2, see Fig. 11.

*Theorem 5.* If a state with positive partial transpose $\rho$ approximates strictly irreducible private dit (pdit) $\|\rho - \gamma_{\langle d_k,d_s \rangle}\| \leqslant \epsilon$ for $\frac{d_k-1}{d_s+d_k(d_k-1)} \leqslant \epsilon < \frac{1}{d_k}$, $\sum_{i \neq j} \|A^\Gamma_{ij,ji}\| \leqslant \epsilon$, and its conditional shield states are separable, then it serves as a two-way $(\theta, \eta)$-good secure network scheme $S_\rho$ with degree $\Delta$, and its overhead is lower bounded with

$$V(S_\rho) \geqslant M(\rho)\left(1 - \frac{\epsilon}{2} - f(d_k, \epsilon)\right), \quad (40)$$

$$f(d_k, \epsilon) := \frac{\log_2 d_k + (1 + \frac{\epsilon}{2})h\left(\frac{\frac{\epsilon}{2}}{1+\frac{\epsilon}{2}}\right)}{\log_2 d_k + \log_2\left(\frac{d_k-1}{\epsilon}\right) + \log_2(1 - \epsilon d_k)}, \quad (41)$$

with $\eta = \log_2 d_k - 8\epsilon \log_2 d_k - 4h(\epsilon)$ [where $h(.)$ is the binary Shannon entropy] and $\theta = 2(\sqrt{\epsilon} + \epsilon) \log_2 \dim_H(\rho) + (1 + 2\sqrt{\epsilon} + 2\epsilon)h(\frac{\sqrt{\epsilon}+\epsilon}{\frac{1}{2}+\sqrt{\epsilon}+\epsilon})$.

For the proof of the above theorem see Appendix. For the lower bound on percentage of memory overhead from this theorem for different values of $d_k$ see Fig. 12, while lower bounds on gap between $\eta$ and $\theta$ for the presented scheme are depicted in Fig. 13.

### A. On relaxation of the honest-but-curious attack assumption

In this section, we discuss in more detail a possible relaxation of the assumption 5 (about the honest-but-curious attack). We argue that one can combine our countermeasure with the protection of the data and/or key of the node.

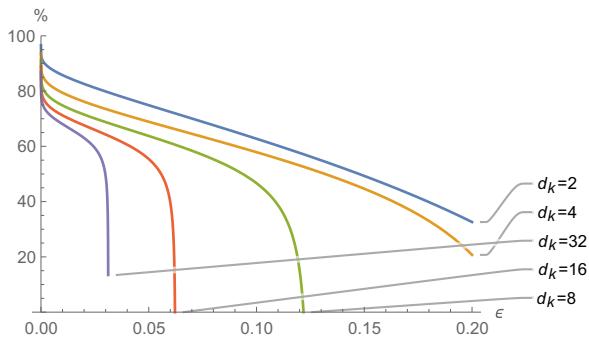

FIG. 12. Plots of lower bound on percentage of memory overhead from theorem 5, with different values of $d_k$.

Indeed, there two major attacks that are in due. The first aims to learn the key of some of the links of the hub with end-users. The second concentrates on the direct learning of classical data stored in the hub's server. One can also consider a mixed strategy aiming at learning both types of data. We note that in order to secure any type of data, one can perform the quantum one-time-pad [34], i.e., rotate each qubit randomly by one of the four Pauli operations. However, this type of encryption does not allow for manipulating the data by the honest party. It only shifts the problem of hacking to the place where the keys of the randomness are stored. A more clever solution involves the so-called *homomorphic encryption* [35,36]. This one aims at allowing to execute some quantum operation on encrypted data without its decryption. Such a solution can be therefore composed with our countermeasure. In particular the efficiency of our scheme (the gap between distillable and repeated key) after composition with the homomorphic encryption stays the same. Indeed, this encryption can be viewed as a local operation on the hub, while our scheme is secure against such operations. In turn, effectiveness against an LOCC protocol (tripartite or from the hub to the users) does not change. However, homomorphic encryption costs a non-negligible amount of quantum memory. In the proposed solution for a number of protected qubits $q$ gets enlarged to $a \times q$ where $a$ is a natural number (constant of the solution). We can still measure the efficiency of such a complex solution with the introduced memory overhead. In this case, it reads

$$V(S'_\rho) = M'(\rho)(1 - D'(\rho)), \quad (42)$$

where $M'(\rho) = \Delta[a(\log_2 \dim_H(\rho))]$, $D'(\rho) = K_D(\rho)/[a(\log_2 \dim_H(\rho))]$. This implies that $V(S'_\rho) - V(S_\rho) = (a - 1)\log_2 \dim_H(\rho)$. The value of $a$ can vary depending on a chosen protocol for homomorphic encryption. In Ref. [36], the value of $a$ is modest, as it equals 3. We note here, that we count only the quantum memory of this solution, rather than classical, as the former is hard to be realized experimentally. In such an approach, the hub has to store classical keys needed for the homomorphic encryption in some separate memory, which is inaccessible to the hacker. However, the quantum part of encryption can be exposed to attacks against the reading of the data.

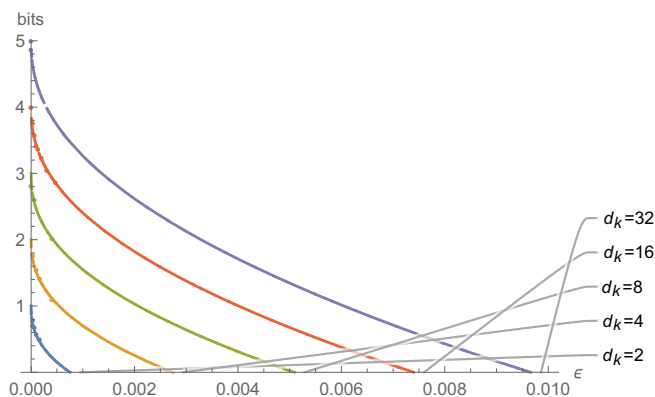

FIG. 13. Plots of lower bounds on gap between $\eta$ and $\theta$ for the scheme in theorem 5. For obtaining possibly optimal value of upper bounds we attribute with $d_s$ its minimal possible value of $d_s = \lceil \frac{d_k - 1 - \epsilon d_k(d_k-1)}{\epsilon} \rceil$.

Finally, let us note that in the above countermeasure, the hacker, in spite of the impossibility of reading, can modify the data. Related countermeasure that can be composed with ours is the intrusion detection obtained recently via the so-called trap-codes [37]. In this case, the memory overhead can also report the cost of relaxation of the security of the data.

## VII. CONCLUSIONS

In this manuscript, we have observed a particular attack on quantum network, and studied the quantum memory cost of its remedy—the hybrid quantum network. A common approach in designing quantum-secured Internet is to connect its nodes via pure entangled states or channels that distribute such entanglement. In this paper, we observe that this practice is not needed for a number of nodes of the Internet, and moreover, would open a threat.

As a case study of such a threat, we consider the possibility of performing entanglement swapping between the data basis of the hub and its two end-users Adam and Eve. We imagine that in future due to development of quantum technologies the link between each of them and the hub would be a quantum one. As a countermeasure, we propose to replace these links with those sharing/distributing bound entangled states which approximate private states. As for end-users, it is enough to communicate only classical information with the hub. What is more a functionality to pass a quantum state seems not only to be a redundant feature but also opens a gateway for a possible abuse.

While in the case of a maximally entangled state, one can generate 1 bit of key per 1 qubit of local memory, this is not the case for mixed entangled quantum states. We, therefore, study the memory cost of the proposed solution. We have introduced two notions: (i) that of a scheme (a choice of states shared by the node and users) and (ii) that of the memory overhead. The latter quantity reports how many qubits of the memory are not directly used up to generate key, but only assures security of its generation. We then focus on schemes that are represented by a single quantum state distributed in all the links. As the quality of the scheme, we propose the gap between the key that can be obtained from the state and the upper bound on the key that can be obtained via hacking. We called it a gap of the scheme.

We first focus on what is more or less straightforward to obtain from the well-established facts in entanglement theory based approach to quantum cryptography. This leads us to two different but asymptotically equivalent lower bounds for the memory overhead of the scheme. One is for private states, and the other for all quantum states. It implies that at least half of the memory of the scheme need to assist security of the scheme rather than can be turned to security itself.

We then consider particular bound entangled states as well as private states for which we know the construction of our proposal can be realized. These are PPT sates that approximate private states and that are at the same time highly indistinguishable by LOPC operations from their attacked versions, which are separable. Although, in general, the overhead, in that case, is asymptotically 1, the convergence to 1 is modest.

The presented results allow to tune the exemplary states to the size of the gap of the scheme. As a byproduct, we have both sharpen the lower bounds on the distance between PPT and private bits and gave the first lower bound on this distance between PPT states and private dits for arbitrary dimension of the key part $d_k$. It would be then interesting to find the schemes based on private dits, rather than those that are based on tensor products of private bits.

Let us note here that we consider the attack to be honest-but curious. Both in the case of quantum repeater and the proposed hybrid repeater, the nodes can be hijacked, and in principle, the data can be traded via blackmail, and therefore, as we have discussed, should be kept e.g., homomorphically encrypted [35,36]. Finding the most effective scheme in terms of memory is an important open problem.

Finally, we admit, that another simple to consider solution for the considered threat, is to live with the fact of possibility of a malware and let every registered user of a node be connected with any other by quantum switch (no matter what is the type of the node) and sell e.g. utility. This, however, would need to be done at a certain price, in similarity to a utility that any smart phone can be turned into a network router within the price of the subscription. In general, one can ask for any other nontransitive property (nonhackable), that can be incorporated to provide security. That will be studied elsewhere (see in this context recent Ref. [38]).

While large effort to make QI happen is begin taken [7], it is also important to know a novel, inherently quantum threats that can come from the new quantum network design. To our knowledge this direction of research needs separate attention, as has not been studied in deep so far [11,12].

## ACKNOWLEDGMENTS

K.H., O.S., and M.W. acknowledge grant Sonata Bis 5 (No. 2015/18/E/ST2/00327) from the National Science Center. K.H. and M.W. acknowledge partial support by the Foundation for Polish Science through IRAP project co-financed by EU within Smart Growth Operational Programme (Contract No. 2018/MAB/5).

## APPENDIX

*Proof of Observation 1.*

$$E_D^{\rightarrow}(\gamma_{d_k,d_s}) \geqslant I_{\text{coh}}(AA'\rangle BB')_{\gamma_{d_k,d_s}} \quad \text{(A1)}$$

$$= S(BB')_{\gamma_{d_k,d_s}} - S(AA'BB')_{\gamma_{d_k,d_s}} \quad \text{(A2)}$$

$$= \log_2 d_k + \sum_i \frac{1}{d_k} S(\sigma_{iB'}) - S(\rho_{A'B'}) \quad \text{(A3)}$$

$$= \log_2 d_k + \sum_i \frac{1}{d_k}[S(\text{Tr}_{A'} U_i \rho_{A'B'} U_i^\dagger) - S(U_i \rho_{A'B'} U_i^\dagger)] \quad \text{(A4)}$$

$$= \log_2 d_k + \sum_i \frac{1}{d_k}[S(B') - S(A'B')]_{[U_i\rho_{A'B'}U_i^\dagger]} \quad \text{(A5)}$$

$$= \log_2 d_k + \sum_i \frac{1}{d_k} I_{\text{coh}}(A'\rangle B')_{[U_i\rho_{A'B'}U_i^\dagger]} \quad \text{(A6)}$$

$$= \log_2 d_k + \sum_i \frac{1}{d_k} I_{\text{coh}}(A'\rangle B')_{\sigma_i}, \quad \text{(A7)}$$

where $\sigma_{iB'} = \text{Tr}_{A'} U_i \rho_{A'B'} U_i^\dagger$.

The first inequality is due to the fact that the one-way distillable entanglement is lower bounded by the coherent information. Then the first equality follows from direct calculation, and the fact that $S(\sum_i \frac{1}{d_k}|i\rangle\langle i| \otimes Tr_{A'} U_i \rho_{A'B'} U_i^\dagger) = \log_2 d_k + \sum_i \frac{1}{d_k} S(Tr_{A'} U_i \rho_{A'B'} U_i^\dagger)$. Equality $S(AA'BB') = S(\rho_{A'B'})$ comes from the construction: the private state is unitarily equivalent to $\psi \otimes \rho$ (where $\psi$ is maximally entangled state of dimension $d_k^2$), and the entropy is invariant under unitary transformations, additive and zero for pure states. In the equality (A4) we add the unitary transformations to $\rho_{A'B'}$ which is assured by mentioned property of entropy: $S(\rho_{A'B'}) = \sum_i \frac{1}{d_k} S(U_i \rho_{A'B'} U_i^\dagger)$. We ten observe that $S(B') - S(A'B')_{U_i \rho_{A'B'} U_i^\dagger}$ is nothing but the coherent information of $\sigma_i$ for each $i$. Hence the final formula involves average value of the coherent information evaluated for states $\sigma_i \equiv U_i \rho_{A'B'} U_i^\dagger$. ∎

*Proof of theorem 1.* Below we present a sequence of inequalities, that altogether allow to prove the theorem.

$$\theta \geqslant R^{\rightarrow}(\gamma_{d_k,d_s}) \geqslant E_D^{\rightarrow}(\gamma_{d_k,d_s}) \geqslant \log_2 d_k +$$
$$\sum_i \frac{1}{d_k} I_{\text{coh}}(A'\rangle B')_{\sigma_i} \geqslant \log_2 d_k - \log_2 d_s, \quad \text{(A8)}$$

The first inequality comes from our assumption that $\gamma_{d_k,d_s}$ is an $(\theta, \log_2 d_k)$-good one-way secure network scheme. The second inequality is supported by the fact, that one can distill $R^{\rightarrow}(\rho)$ singlets and use them for teleportation. One of methods to repeat key is to distill $E_D^{\rightarrow}$ of pure entanglement between H and A and H and B, respectively. This is followed by entanglement swapping protocol [8]. The third inequality comes from Eq. (14). The final inequality is due to corollary 1.

Thanks to the above inequality (A8), we can upper bound the density of the private key as follows:

$$D(\gamma_{d_k,d_s}) = \frac{\log_2 d_k}{\log_2 \dim_H} = \frac{\log_2 d_k}{\log_2 d_k + \log_2 d_s} \quad (A9)$$

$$\leqslant \frac{\log_2 d_k}{2\log_2 d_k - \theta} = \frac{1}{2 - \frac{\theta}{\log_{2_{d_k}}}}. \quad (A10)$$

From Eq. (11), we have

$$V(S_{\gamma_{d_k,d_s}}) \geqslant M(\gamma_{d_k,d_s})\left(1 - \frac{1}{2 - \frac{\theta}{\log_2 d_k}}\right) \quad (A11)$$

$$\approx M(\gamma_{d_k,d_s})\left(\frac{1}{2} - \frac{\theta}{4\log_2 d_k} + O(\theta^2)\right) \approx_{\theta \approx 0} \frac{1}{2}M(\gamma_{d_k,d_s}), \quad (A12)$$

what ends the proof. ∎

*Proof of theorem 2.* Because $\theta \geqslant R^{\leftrightarrow}(\rho) \geqslant E_D(\rho) \geqslant I_{\text{coh}}(H \rangle A)$, and it has nonpositive coherent information $I_{\text{coh}}(H \rangle A)$ [27], thus distillable key has to fulfill:

$$K_D(\rho) \leqslant E_{\text{sq}}(\rho) \leqslant \tfrac{1}{2}I(\rho) = \tfrac{1}{2}S(H) + \tfrac{1}{2}I_{\text{coh}}(H \rangle A)$$

$$\leqslant \tfrac{1}{2}(\log_2 d_k + \log_2 d_s) + \theta, \quad (A13)$$

where $E_{sq}(\rho_{A,B}) = \inf_{\rho^{ABE} \in S_{\text{Ext}}} \tfrac{1}{2}I(A;B|E)$ is the squashed entanglement [39], and the next inequality is by the definition of $E_{\text{sq}}$. Owing to the fact that $K_D(\rho) \geqslant \eta$, we obtain

$$\eta \leqslant \tfrac{1}{2}(\log_2 d_k + \log_2 d_s) + \theta, \quad (A14)$$

$$\mathcal{D}(\rho) \leqslant \frac{\eta}{\log_2 d_H} \leqslant \frac{\tfrac{1}{2}\log_2 d_H + \theta}{\log_2 d_H} = \frac{1}{2} + \frac{\theta}{\log_2 d_H}, \quad (A15)$$

$$V(S_\rho) \geqslant M(\rho)\left(\frac{1}{2} - \frac{\theta}{\log_2 d_H}\right) \approx_{\theta \approx 0} \tfrac{1}{2}M(\rho). \quad (A16)$$

∎

*Proof of observation 2.* The first inequality comes from the result of Christandl and Ferrara [19]. There is

$$R^{\rightarrow}(\gamma) \leqslant E_D^{\rightarrow}(\gamma \otimes \gamma). \quad (A17)$$

The distillable entanglement is upper bounded by the log-negativity:

$$E_D^{\rightarrow}(\gamma \otimes \gamma) \leqslant \log_2 \|\gamma^\Gamma \otimes \gamma^\Gamma\| = 2\log_2 \|\gamma^\Gamma\|, \quad (A18)$$

where equality comes from the additivity of the log-negativity. We upper bound log-negativity as follows:

$$\|\gamma^\Gamma\| = \left\| \frac{1}{d_k}\sum_{ij}|ij\rangle\langle ji| \otimes X_{ij}^\Gamma \right\| \quad (A19)$$

$$= \left\| \frac{1}{d_k}\sum_i |ii\rangle\langle ii| \otimes X_{ii}^\Gamma + \frac{1}{d_k}\sum_{i \neq j}|ij\rangle\langle ji| \otimes X_{ij}^\Gamma \right\| \quad (A20)$$

$$\leqslant \left\| \frac{1}{d_k}\sum_i |ii\rangle\langle ii| \otimes X_{ii}^\Gamma \right\| + \left\| \frac{1}{d_k}\sum_{i \neq j}|ij\rangle\langle ji| \otimes X_{ij}^\Gamma \right\| \quad (A21)$$

$$= 1 + \|\gamma^\Gamma - \hat{\gamma}^\Gamma\|. \quad (A22)$$

The last equality is obtained due to the fact that $X_{ii} \in PPT$. Finally, because logarithm is strictly increasing, we have $2\log_2 \|\gamma^\Gamma\| \leqslant 2\log_2(1 + \|\gamma^\Gamma - \hat{\gamma}^\Gamma\|)$, and hence

$$E_D^{\rightarrow}(\gamma \otimes \gamma) \leqslant 2\log_2(1 + \|\gamma^\Gamma - \hat{\gamma}^\Gamma\|). \quad (A23)$$

This implies by virtue of Eq. (A17):

$$R^{\rightarrow}(\gamma) \leqslant 2\log_2(1 + \|\gamma^\Gamma - \hat{\gamma}^\Gamma\|). \quad (A24)$$

∎

*Proof of observation 3.* By direct calculations we have

$$\|\gamma^\Gamma - \hat{\gamma}^\Gamma\| \quad (A25)$$

$$= \left\| \sum_{i,j} \frac{1}{d_k}|ij\rangle\langle ij| \otimes X_{ij}^\Gamma - \sum_i \frac{1}{d_k}|ii\rangle\langle ii| \otimes X_{ii}^\Gamma \right\| \quad (A26)$$

$$= \left\| \sum_{i \neq j} \frac{1}{d_k}|ij\rangle\langle ij| \otimes X_{ij}^\Gamma \right\| \quad (A27)$$

$$= \sum_{i \neq j} \frac{1}{d_k}\|X_{ij}^\Gamma\|. \quad (A28)$$

∎

*Proof of lemma 1.* A pdit $\gamma_{d_k,d_s}$ has $d_k^2 - d_k$ off-diagonal block elements $X_{ij}$, and $\|X_{ij}\| \geqslant 0$. From observation 3 we have that $\sum_{i \neq j} \frac{1}{d_k}\|X_{ij}^\Gamma\| \leqslant \epsilon$, for some small $\epsilon \geqslant \|\gamma_{d_k,d_s}^\Gamma - \hat{\gamma}_{d_k,d_s}^\Gamma\|$. Then among those block elements there clearly has to be a one such that $\frac{1}{d_k}\|X_{i_0,j_0}^\Gamma\| \leqslant \frac{\epsilon}{d_k^2 - d_k}$ as a property of mean value, hence

$$\|X_{i_0,j_0}^\Gamma\| \leqslant \frac{\epsilon d_k}{d_k^2 - d_k} = \frac{\epsilon}{d_k - 1}. \quad (A29)$$

We know from [23], that $\|X_{ij}\| \leqslant d_s\|X_{ij}^\Gamma\|$. Hence, for arbitrary $i$ and $j$, we have

$$d_s\|X_{ij}^\Gamma\| \geqslant \|X_{ij}\| = \|U_i \sigma U_j^\dagger\| = \|\sigma\| = 1. \quad (A30)$$

In particular $1 \leqslant d_s\|X_{i_0,j_0}^\Gamma\|$. Then $1 \leqslant d_s \frac{\epsilon}{d_k-1}$ and finally $d_s \geqslant \frac{d_k-1}{\epsilon}$. ∎

*Proof of theorem 3.* From observation 2: $R^{\rightarrow}(\gamma_{\langle d_k,d_s \rangle}) \leqslant 2\log_2(1 + \epsilon)$. Further from irreducability of $\gamma_{\langle d_k,d_s \rangle}$, we have that $K_D(\gamma_{\langle d_k,d_s \rangle}) = \log_2 d_k$, and the lower bound for $V(S_{\gamma_{\langle d_k,d_s \rangle}})$ we obtain in the following way

$$\mathcal{D}(\gamma_{\langle d_k,d_s \rangle}) = \frac{K_D(\gamma_{\langle d_k,d_s \rangle})}{\log_2 d_k + \log_2 d_s}$$

$$= \frac{\log_2 d_k}{\log_2 d_k + \log_2 d_s} \leqslant \frac{\log_2 d_k}{\log_2 d_k + \log_2 \frac{d_k-1}{\epsilon}}. \quad (A31)$$

Thus

$$V(S_{\gamma_{\langle d_k,d_s \rangle}}) \geqslant M\gamma_{\langle d_k,d_s \rangle}\left(1 - \frac{\log_2 d_k}{\log_2 d_k + \log_2 \frac{d_k-1}{\epsilon}}\right)$$

$$\approx_{\epsilon=0} M(\gamma_{\langle d_k,d_s \rangle}), \quad (A32)$$

where the first inequality is a consequence of lemma 1. ∎

*Proof of proposition 1.* In this proof, partial transposition $\Gamma$ and the operation of diag(·) are assumed to be evaluated in computational basis. Furthermore we assume $\|A^\Gamma_{01,10}\| \leqslant \epsilon$. Using the results in Ref. [23], we know

$$||\rho - \gamma_{\langle 2, d_s \rangle}|| \leqslant \epsilon \Rightarrow ||A^\Gamma_{0011}|| \leqslant \epsilon. \quad (A33)$$

We define a projection

$$\Pi := (|00\rangle\langle 00| + |11\rangle\langle 11|) \otimes I_{d_s^2}. \quad (A34)$$

Notice that $\Pi \gamma_{\langle 2, d_s \rangle} \Pi = \gamma_{\langle 2, d_s \rangle}$, and let us define subnormalized state

$$\rho^\Gamma_\Pi := \Pi \rho^\Gamma \Pi. \quad (A35)$$

From one of assumptions, we have

$$\|\rho^\Gamma_\Pi - \mathrm{diag}(\rho^\Gamma_\Pi)\| = 2\|A^\Gamma_{01,10}\| \leqslant 2\epsilon, \quad (A36)$$

where $\mathrm{diag}(\cdot)$ refers to an operation that projects the key part to its diagonal, i.e, it acts in the following way:

$$\mathrm{diag}\left(\sum_{i,j,k,l} |ij\rangle\langle kl| \otimes A_{ij,kl}\right) := \sum_{i,j,} |ij\rangle\langle ij| \otimes A_{ij,ij}. \quad (A37)$$

We define a CPTP operation $\phi$ and corresponding Kraus operators:

$$\phi(\rho) := \sum_{i=1}^{2} K_i \rho K_i^\dagger, \quad (A38)$$

$$K_1 := |01\rangle\langle 01| \otimes I_{d_s^2}, \quad (A39)$$

$$K_2 := |10\rangle\langle 10| \otimes I_{d_s^2}. \quad (A40)$$

We employ the above to upper bound some traces of certain diagonal block of $\rho^\Gamma$. Since the trace norm is nonincreasing under CPTP maps and for $i \neq j$, we have $X_{ij,ij} = 0$:

$$\epsilon \geqslant \|\rho - \gamma_{\langle 2, d_s \rangle}\| \geqslant \|\phi(\rho) - \phi(\gamma_{\langle 2, d_s \rangle})\| \quad (A41)$$

$$= \||01\rangle\langle 01| \otimes A_{01,01} + |10\rangle\langle 10| \otimes A_{10,10}\| \quad (A42)$$

$$= \|A_{01,01}\| + \|A_{10,10}\| = \mathrm{Tr} A_{01,01} + \mathrm{Tr} A_{10,10} \quad (A43)$$

$$= \mathrm{Tr} A^\Gamma_{01,01} + \mathrm{Tr} A^\Gamma_{10,10}, \quad (A44)$$

where we used a property that the trace of hermitean positive semidefinite matrix is invariant under partial transpose. We use now Eqs. (A41)–(A44) to lower bound the following quantity:

$$\mathrm{Tr}(\Pi \rho^\Gamma \Pi) = \mathrm{Tr} A^\Gamma_{00,00} + \mathrm{Tr} A^\Gamma_{11,11} \quad (A45)$$

$$= 1 - \mathrm{Tr} A^\Gamma_{01,01} + \mathrm{Tr} A^\Gamma_{10,10} \geqslant 1 - \epsilon. \quad (A46)$$

As a byproduct notice that

$$\|\rho^\Gamma_\Pi\| = \mathrm{Tr}(\rho^\Gamma_\Pi) \equiv \mathrm{Tr}(\Pi \rho^\Gamma \Pi) \geqslant 1 - \epsilon. \quad (A47)$$

We employ now the "gentle measurement lemma" [40–42], saying that for all positive semidefinite operators $\sigma$, and $0 \leqslant H \leqslant 1$, one has

$$\|\sigma - \sqrt{H}\sigma\sqrt{H}\| \leqslant 2\sqrt{\mathrm{Tr}(\sigma)}\sqrt{\mathrm{Tr}(\sigma(I - H))}. \quad (A48)$$

Since $\Pi$ is a projector, and $\rho^\Gamma$ is normalized, from Eqs. (A45),(A46), and (A48) we find

$$\|\rho^\Gamma - \rho^\Gamma_\Pi\| \leqslant 2\sqrt{1 - \mathrm{Tr}(\Pi \rho^\Gamma \Pi)} \leqslant 2\sqrt{\epsilon}, \quad (A49)$$

where we used cyclic property of the trace. Using the triangle inequality twice, the fact that $\|\rho^\Gamma_\Pi\| \equiv \|\mathrm{diag}(\rho^\Gamma_\Pi)\|$, and inequalities in (A36), (A47), and (A49), we obtain

$$\left\|\rho^\Gamma - \frac{\mathrm{diag}(\rho^\Gamma_\Pi)}{\|\rho^\Gamma_\Pi\|}\right\| \leqslant \|\rho^\Gamma - \mathrm{diag}(\rho^\Gamma_\Pi)\| \quad (A50)$$

$$+ \left\|\mathrm{diag}(\rho^\Gamma_\Pi) - \frac{\mathrm{diag}(\rho^\Gamma_\Pi)}{\|\rho^\Gamma_\Pi\|}\right\| \quad (A51)$$

$$= \|\rho^\Gamma - \rho^\Gamma_\Pi + (\rho^\Gamma_\Pi - \mathrm{diag}(\rho^\Gamma_\Pi))\| + (1 - \|\rho^\Gamma_\Pi\|) \quad (A52)$$

$$\leqslant \|\rho^\Gamma - \rho^\Gamma_\Pi\| + \|\rho^\Gamma_\Pi - \mathrm{diag}(\rho^\Gamma_\Pi)\| + \epsilon \quad (A53)$$

$$\leqslant 2\sqrt{\epsilon} + 2\epsilon + \epsilon = 2\left(\sqrt{\epsilon} + \frac{3}{2}\epsilon\right). \quad (A54)$$

From the Refs. [15,22], two-way repeater rate is upper bounded in the following way:

$$R^\leftrightarrow(\rho) \leqslant K_D(\rho^\Gamma) \leqslant E_r(\rho^\Gamma). \quad (A55)$$

While employing asymptotic continuity of the relative entropy of entanglement $E_r$ [43,44], we obtain

$$\left|E_r(\rho^\Gamma) - E_r\left(\frac{\mathrm{diag}(\rho^\Gamma_\Pi)}{\|\rho^\Gamma_\Pi\|}\right)\right| \leqslant \xi \log_2 \dim_H(\rho^\Gamma)$$

$$+ (1 + \xi) h\left(\frac{\xi}{1 + \xi}\right) \quad (A56)$$

$$\Rightarrow E_r(\rho^\Gamma) \leqslant E_r\left(\frac{\mathrm{diag}(\rho^\Gamma_\Pi)}{\|\rho^\Gamma_\Pi\|}\right)$$

$$+ \xi \log_2 \dim_H(\rho^\Gamma) + (1 + \xi) h\left(\frac{\xi}{1 + \xi}\right), \quad (A57)$$

where $\xi = 2(\sqrt{\epsilon} + \frac{3}{2}\epsilon)$. From Eq. (A55), we have then

$$R^\leftrightarrow(\rho) \leqslant E_r\left(\frac{\mathrm{diag}(\rho^\Gamma_\Pi)}{\|\rho^\Gamma_\Pi\|}\right) + \xi \log_2 \dim_H(\rho^\Gamma) \quad (A58)$$

$$+ (1 + \xi) h\left(\frac{\xi}{1 + \xi}\right). \quad (A59)$$

Blocks of $\mathrm{diag}(\rho^\Gamma)$ are separable by assumption. Since nonzero blocks of $\mathrm{diag}(\rho^\Gamma_\Pi)$ are identical to corresponding blocks of $\mathrm{diag}(\rho^\Gamma)$ they are also separable. This implies that the relative entropy of entanglement of $\frac{\mathrm{diag}(\rho^\Gamma_\Pi)}{\|\rho^\Gamma_\Pi\|}$, from its definition reads 0. Knowing that $d_k = 2$ and that dimension of matrix is invariant under partial transpose, we obtain an upper bound.

$$R^\leftrightarrow(\rho) \leqslant 2\left(\sqrt{\epsilon} + \frac{3}{2}\epsilon\right)(1 + \log_2 d_s) \quad (A60)$$

$$+ (1 + 2\sqrt{\epsilon} + 3\epsilon) h\left(\frac{2\sqrt{\epsilon} + 3\epsilon}{1 + 2\sqrt{\epsilon} + 3\epsilon}\right). \quad (A61)$$

∎

*Proof of theorem 4.* We work under an assumption that $\frac{1}{2(d_s+1)} \leqslant ||\rho - \gamma_{\langle 2, d_s \rangle}|| \leqslant \epsilon < \frac{1}{2}$. The first step is to upper bound key rate with relative entropy (see Ref. [22]):

$$K_D(\rho) \leqslant E_r(\rho). \quad (A62)$$

Then we make use of asymptotic continuity of quantum relative entropy [43,44].

$$\left| E_r(\rho) - E_r(\gamma_{\langle 2, d_s \rangle}) \right| \leqslant \frac{\epsilon}{2} \log_2 \dim_H(\rho) + \left(1 + \frac{\epsilon}{2}\right) h\left(\frac{\frac{\epsilon}{2}}{1 + \frac{\epsilon}{2}}\right) \quad \text{(A63)}$$

$$\Rightarrow E_r(\rho) \leqslant E_r(\gamma_{\langle 2, d_s \rangle}) + \frac{\epsilon}{2} \log_2 \dim_H(\rho) + \left(1 + \frac{\epsilon}{2}\right) h\left(\frac{\frac{\epsilon}{2}}{1 + \frac{\epsilon}{2}}\right). \quad \text{(A64)}$$

Since $E_r(\gamma_{\langle d_k, d_s \rangle}) \leqslant \log_2 d_k$ [22], by combining Eqs. (A62) and (A64), we have

$$K_D(\rho) \leqslant \log_2 d_k + \frac{\epsilon}{2} \log_2 \dim_H \rho + \left(1 + \frac{\epsilon}{2}\right) h\left(\frac{\frac{\epsilon}{2}}{1 + \frac{\epsilon}{2}}\right). \quad \text{(A65)}$$

From Ref. [23], we know that $||\rho - \gamma_{\langle 2, d_s \rangle}|| \geqslant \frac{1}{2(d_s+1)}$, what together with the initial condition ($\epsilon < \frac{1}{2}$) yields

$$\log_2 d_s \geqslant \log_2 \left(\frac{1 - 2\epsilon}{2\epsilon}\right). \quad \text{(A66)}$$

The overhead of the scheme is then lower bounded

$$V(\rho) = M(\rho)\left(1 - \frac{K_D^{\rightarrow}(\rho)}{\log_2 \dim_H(\rho)}\right) \quad \text{(A67)}$$

$$\geqslant M(\rho)\left(1 - \frac{\log_2 d_k + \frac{\epsilon}{2} \log_2 \dim_H(\rho) + \left(1 + \frac{\epsilon}{2}\right) h\left(\frac{\frac{\epsilon}{2}}{1 + \frac{\epsilon}{2}}\right)}{\log_2 \dim_H(\rho)}\right) \quad \text{(A68)}$$

$$\geqslant M(\rho)\left(1 - \frac{\log_2 d_k + \left(1 + \frac{\epsilon}{2}\right) h\left(\frac{\frac{\epsilon}{2}}{1 + \frac{\epsilon}{2}}\right)}{\log_2 d_k + \log_2\left(\frac{1 - 2\epsilon}{2\epsilon}\right)} - \frac{\epsilon}{2}\right) \quad \text{(A69)}$$

$$= M(\rho)\left(1 - \frac{1 + \left(1 + \frac{\epsilon}{2}\right) h\left(\frac{\frac{\epsilon}{2}}{1 + \frac{\epsilon}{2}}\right)}{1 + \log_2\left(\frac{1 - 2\epsilon}{2\epsilon}\right)} - \frac{\epsilon}{2}\right). \quad \text{(A70)}$$

Where we used $\dim_H(\rho) = \dim_H(\gamma_{\langle 2, d_s \rangle})$ and $d_k = 2$.

Now we have to design an appropriate lower bound on $K_D$. Following arguments of Ref. [45], the operation of privacy squeezing does not increase the trace distance $||\rho^{\text{ps}} - \gamma^{\text{ps}}_{\langle 2, d_s \rangle}|| \leqslant \epsilon$. Moreover after this operation private state (strictly irreducible in this case) turns into one of the two Bell states $\gamma^{\text{ps}}_{\langle 2, d_s \rangle} \equiv \psi$. In general, the following inequalities hold:

$$K_D^{\rightarrow}(\rho^{\text{ps}}) \leqslant K_D^{\rightarrow}(\rho) \leqslant K_D(\rho). \quad \text{(A71)}$$

On the other hand due to lemma V.3. in Ref. [46], both one-way and two-way key rates are lower bounded with

$$1 - 8\epsilon \log_2 \dim_H \left(\gamma^{\text{ps}}_{\langle 2, d_s \rangle}\right) - 4h(\epsilon) \leqslant K_D^{\rightarrow}(\rho^{\text{ps}}). \quad \text{(A72)}$$

From Eqs. (A71) and (A72), and the fact $\dim_H(\rho^{\text{ps}}) = 2$, we obtain

$$K_D(\rho) \geqslant \eta := 1 - 8\epsilon - 4h(\epsilon). \quad \text{(A73)}$$

Form proposition 1, the rate of the repeater is upper bounded with

$$R^{\leftrightarrow}(\rho) \leqslant \theta := 2\left(\sqrt{\epsilon} + \frac{3}{2}\epsilon\right)(1 + \log_2 d_s) \quad \text{(A74)}$$

$$+ (1 + 2\sqrt{\epsilon} + 3\epsilon) h\left(\frac{2\sqrt{\epsilon} + 3\epsilon}{1 + 2\sqrt{\epsilon} + 3\epsilon}\right). \quad \text{(A75)}$$

∎

*Notation 6.* We denote projectors $P_{i,j}$ and $P_i$ for $i \neq j$

$$P_{i,j} = |ii\rangle\langle ii| + |jj\rangle\langle jj|, \quad \text{(A76)}$$

$$P_i = |ii\rangle\langle ii|. \quad \text{(A77)}$$

The following identities hold.
*Fact 1.* We have the following identities:

$$(P_{i,j} \otimes I)\left(\sum_{ijkl} |ij\rangle\langle kl| \otimes A_{ij,kl}\right)(P_{i,j} \otimes I) \quad \text{(A78)}$$

$$= |ii\rangle\langle ii| \otimes A_{ii,ii} + |ii\rangle\langle jj| \otimes A_{ii,jj} \quad \text{(A79)}$$

$$+ |jj\rangle\langle ii| \otimes A_{jj,ii} + |jj\rangle\langle jj| \otimes A_{jj,jj} \quad \text{(A80)}$$

and also

$$(P_i \otimes I)\left(\sum_{ijkl} |ij\rangle\langle kl| \otimes A_{ij,kl}\right)(P_i \otimes I) \quad \text{(A81)}$$

$$= |ii\rangle\langle ii| \otimes A_{ii,ii}. \quad \text{(A82)}$$

*Notation 7.* For the proofs of observation 4 we abuse the notation denoting $\frac{1}{d_k} X_{ii,jj} \to X_{ii,jj}$, $P_{i,j} \otimes I \to P_{i,j}$, and $P_i \otimes I \to P_i$ for conciseness.

*Proof of observation 4.* We start with proving first inequality (34). Using the contractivity of the trace norm, we have

$$\|\rho - \gamma\| \leqslant \epsilon \Rightarrow \|P_{i,j} \rho P_{i,j} - P_{i,j} \gamma P_{i,j}\| \leqslant \epsilon. \quad \text{(A83)}$$

Thus

$$\| |ii\rangle\langle ii| \otimes (A_{ii,ii} - X_{ii,ii}) \quad \text{(A84)}$$

$$+ |ii\rangle\langle jj| \otimes (A_{ii,jj} - X_{ii,jj}) \quad \text{(A85)}$$

$$+ |jj\rangle\langle ii| \otimes (A_{jj,ii} - X_{jj,ii})$$

$$+ |jj\rangle\langle jj| \otimes (A_{jj,jj} - X_{jj,jj})\| \leqslant \epsilon. \quad \text{(A86)}$$

Using again the norm contractivity property and projector $P_i$, we have

$$\epsilon \geqslant \|\rho - \gamma\| \geqslant \|P_i \rho P_i - P_i \gamma P_i\| \quad \text{(A87)}$$

$$= \sum_i \|A_{ii,ii} - X_{ii,ii}\| \geqslant \|A_{ii,ii} - X_{ii,ii}\|. \quad \text{(A88)}$$

Now we want to prove that

$$\|A_{ii,jj} - X_{ii,jj}\| \leqslant \epsilon. \quad \text{(A89)}$$

Let us express the matrix from LHS of (A86) as follows:

$$M = D + \hat{A}, \quad \text{(A90)}$$

where $M$ is a matrix, $D$ are diagonal elements and $\hat{A}$ are antidiagonal elements. Note that $\|D\| \leqslant \epsilon$ as $\|M\| \leqslant \epsilon$.

We get then

$$\|M\| = \|D + \hat{A}\| \geqslant |\|D\| - \|\hat{A}\|| \quad \text{(A91)}$$
$$\Rightarrow \|\hat{A}\| \leqslant \|M\| + \|D\| \leqslant \|M\| + \epsilon \leqslant 2\epsilon. \quad \text{(A92)}$$

We note then that

$$\|\hat{A}\| = \left\|\begin{matrix} 0 & \hat{A}_{ii,jj} \\ \hat{A}_{ii,jj}^\dagger & 0 \end{matrix}\right\| = 2\|\hat{A}_{ii,jj}\|,$$

hence

$$\|\hat{A}\| = 2\|A_{ii,jj} - X_{ii,jj}\| \leqslant 2\epsilon, \quad \text{(A93)}$$
$$\|A_{ii,jj} - X_{ii,jj}\| \leqslant \epsilon. \quad \text{(A94)}$$

Finally, applying the reverse triangle inequality to Eq. (A94) and having $\|X_{ii,jj}\| = \frac{1}{d_k}$,

$$|\|A_{ii,jj}\| - \frac{1}{d_k}| \leqslant \epsilon \Rightarrow \|A_{ii,jj}\| \geqslant \frac{1}{d_k} - \epsilon. \quad \text{(A95)}$$

Now we prove the second inequality (35). Consider an incomplete von Neumann measurement

$$\{K_{ij}\} = \{|ij\rangle\langle ij| \otimes I\}. \quad \text{(A96)}$$

Using $\|\rho - \gamma\| \leqslant \epsilon$ and contractivity of norm, we obtain

$$\left\|\sum_{ij} K_{ij}\rho K_{ij}^\dagger - \sum_{ij} K_{ij}\gamma K_{ij}^\dagger\right\| \leqslant \epsilon, \quad \text{(A97)}$$

$$\left\|\sum_{ij} |ij\rangle\langle ij| \otimes A_{ij,ij} - \sum_{i} |ii\rangle\langle ii| \otimes X_{ii,ii}\right\| \leqslant \epsilon. \quad \text{(A98)}$$

For $i \neq j$ let $X_{ij,ij} = 0$, then

$$\left\|\sum_{ij} |ij\rangle\langle ij| \otimes A_{ij,ij} - \sum_{ij} |ij\rangle\langle ij| \otimes X_{ij,ij}\right\| \leqslant \epsilon, \quad \text{(A99)}$$

$$\sum_{ij} \||ij\rangle\langle ij| \otimes (A_{ij,ij} - X_{ij,ij})\| \leqslant \epsilon. \quad \text{(A100)}$$

$$\sum_{i\neq j} \|A_{ij,ij} - X_{ij,ij}\| + \sum_{i} \|A_{ii,ii} - X_{ii,ii}\| \leqslant \epsilon. \quad \text{(A101)}$$

Employing the aforementioned condition that $X_{ij,ij}$ vanish and non-negativity of the trace norm, we obtain

$$\sum_{i\neq j} \|A_{ij,ij}\| \leqslant \epsilon. \quad \text{(A102)}$$

∎

*Proof of lemma 2.* We know that $\rho^\Gamma \geqslant 0$. Firstly we construct, a projector on certain $2 \times d_s$ dimensional subspace of $\rho^\Gamma \geqslant 0$.

$$\Pi_0 = (|ij\rangle\langle ij| + |ji\rangle\langle ji|) \otimes I_{d_s^2}, \quad i \neq j. \quad \text{(A103)}$$

Having in mind that $\rho = \sum_{ijkl} |ij\rangle\langle kl| \otimes A_{ij,kl}$, we perform the projection and obtain

$$\Pi_0 \rho^\Gamma \Pi_0 = \begin{bmatrix} A_{ij,ij}^\Gamma & 0 & 0 & A_{ii,jj}^\Gamma \\ 0 & 0 & 0 & 0 \\ 0 & 0 & 0 & 0 \\ A_{ii,jj}^{\Gamma\dagger} & 0 & 0 & A_{ji,ji}^\Gamma \end{bmatrix} \geqslant 0, \quad \text{(A104)}$$

where we used that $A_{jj,ii}^\Gamma = (A_{ii,jj}^\Gamma)^\dagger$, what is a consequence of $\rho^\Gamma$ being Hermitian. Indeed $\Pi_0\rho^\Gamma\Pi_0$ is positive semidefinite since $\Pi_0$ is a Kraus operator. In what follows, we construct a unitary transformation based on singular value decomposition of $A_{ii,jj}^\Gamma = S\Sigma V$.

$$U = |ij\rangle\langle ij| \otimes S^\dagger + |ji\rangle\langle ji| \otimes V. \quad \text{(A105)}$$

Note that $\text{Tr}\Sigma = \|A_{ii,jj}^\Gamma\|$. In the next step, we perform a specific privacy squeezing operation on $\rho^\Gamma$:

$$\rho_{\text{ps.}}^\Gamma = \text{Tr}_{A'B'} U \Pi_0 \rho^\Gamma \Pi_0 U^\dagger. \quad \text{(A106)}$$

What yields following form of a privacy squeezed matrix, which is positive semidefinite,

$$\rho_{\text{ps.}}^\Gamma = \begin{bmatrix} \|A_{ij,ij}\| & 0 & 0 & \|A_{ii,jj}^\Gamma\| \\ 0 & 0 & 0 & 0 \\ 0 & 0 & 0 & 0 \\ \|A_{ii,jj}^\Gamma\| & 0 & 0 & \|A_{ji,ji}\| \end{bmatrix} \geqslant 0. \quad \text{(A107)}$$

Where we used a property of diagonal blocks $\|A_{ij,ij}\| = \text{Tr}A_{ij,ij} = \text{Tr}A_{ij,ij}^\Gamma = \|A_{ij,ij}^\Gamma\|$. Using a basic fact known for positive matrices we have the following dependence between its elements:

$$\|A_{ii,jj}^\Gamma\| \leqslant \frac{\|A_{ij,ij}\| + \|A_{ji,ji}\|}{2}. \quad \text{(A108)}$$

Now we are going to use observation 4. Since the smallest component of the sum is always smaller than an average, we have

$$2\epsilon \geqslant 2\sum_{i\neq j} \|A_{ij,ij}\| = \sum_{i\neq j}(\|A_{ij,ij}\| + \|A_{ji,ji}\|) \quad \text{(A109)}$$

$$= d_k(d_k - 1) \sum_{i\neq j} \frac{(\|A_{ij,ij}\| + \|A_{ji,ji}\|)}{d_k(d_k - 1)} \quad \text{(A110)}$$

$$\geqslant d_k(d_k - 1) \min_{i\neq j}(\|A_{ij,ij}\| + \|A_{ji,ji}\|). \quad \text{(A111)}$$

Since Eq. (A108) is true for all $i \neq j$, we use the smallest element denoted with $i_0 \neq j_0$. Hence form (A108),

$$\|A_{i_0i_0,j_0j_0}^\Gamma\| \leqslant \frac{\epsilon}{d_k(d_k - 1)}. \quad \text{(A112)}$$

By observation 4, $\forall_{i\neq j}$ we have $\|A_{ii,jj}\| \geqslant \frac{1}{d_k} - \epsilon$ and $\|A_{i_0i_0,j_0j_0}^\Gamma\| \leqslant \frac{\epsilon}{d_k^2 - d_k}$. Owing to the fact that under partial transposition the trace norm can not increase by more than the dimension of the matrix (here $d_s$) [23], we have

$$\frac{1}{d_k} - \epsilon \leqslant \|A_{i_0i_0,j_0j_0}\| \leqslant d_s \|A_{i_0i_0,j_0j_0}^\Gamma\| \leqslant \frac{\epsilon}{d_k^2 - d_k} d_s, \quad \text{(A113)}$$

thus,

$$1 - \epsilon d_k \leqslant \frac{d_k \epsilon}{d_k^2 - d_k} d_s = \frac{\epsilon}{d_k - 1} d_s, \quad \text{(A114)}$$

and finally,

$$d_s \geqslant \left(\frac{d_k - 1}{\epsilon}\right)(1 - \epsilon d_k). \quad \text{(A115)}$$

∎

*Proof of corollary 2.* The proof is straightforward consequence of lemma 2. Since the implication stated in Lemma

2 is true for any $\epsilon$ that $\epsilon \geqslant \|\rho - \gamma\|$, we denote with $\epsilon_0$ the one that saturates it. We have the following implication:

$$\|\rho - \gamma_{d_k,d_s}\| = \epsilon_0 \Rightarrow d_s \geqslant \left(\frac{d_k - 1}{\epsilon_0}\right)(1 - \epsilon_0 d_k). \quad (A116)$$

We immediately obtain the following lower bound:

$$\|\rho - \gamma_{d_k,d_s}\| \geqslant \frac{d_k - 1}{d_s + d_k(d_k - 1)}. \quad (A117)$$

∎

*Proof of corollary 3.* We notice that equation (A112) is true also for $d_k = 2$. Since in this dimension there is only a single choice of $i_0 \neq j_0$ (up to Hermitian conjugate), we have:

$$\|A^\Gamma_{00,11}\| \leqslant \frac{\epsilon}{d_k(d_k - 1)} = \frac{\epsilon}{2}. \quad (A118)$$

∎

*Proof of proposition 2.* This proof follows the same steps as the proof of proposition 1. Partial transposition $\Gamma$ and the operation of $\mathrm{diag}(\cdot)$ are assumed to be evaluated in computational basis. Futhermore we assume that $\sum_{i \neq j} \|A^\Gamma_{ij,ji}\| \leqslant \epsilon$. We work under an assumption that $\frac{d_k - 1}{d_s + d_k(d_k - 1)} \leqslant \|\rho - \gamma_{\langle d_k,d_s \rangle}\| \leqslant \epsilon < \frac{1}{d_k}$.

We define a projection and subnormalized state $\rho^\Gamma_\Pi$,

$$\Pi := \sum_{i=0}^{d_k - 1} |ii\rangle\langle ii| \otimes I_{d_s^2}, \quad (A119)$$

$$\rho^\Gamma_\Pi := \Pi \rho^\Gamma \Pi. \quad (A120)$$

We notice then that

$$\|\rho^\Gamma_\Pi - \mathrm{diag}(\rho^\Gamma_\Pi)\| = \sum_{i \neq j} \|A^\Gamma_{ij,ji}\| \leqslant \epsilon, \quad (A121)$$

Where operation of $\mathrm{diag}(\cdot)$ is defined in Eq. (A37).

We anticipate now and calculate the following quantity using equation (A102) again:

$$\mathrm{Tr}(\Pi \rho^\Gamma \Pi) = \sum_{i=0}^{d_k - 1} A^\Gamma_{ii,ii} \quad (A122)$$

$$= \sum_{i=0}^{d_k - 1} A_{ii,ii} = 1 - \sum_{i \neq j} A_{ij,ij} \geqslant 1 - \epsilon. \quad (A123)$$

As a byproduct we notice that

$$\|\rho^\Gamma_\Pi\| = \mathrm{Tr}(\rho^\Gamma_\Pi) = \mathrm{Tr}(\Pi \rho^\Gamma \Pi) \geqslant 1 - \epsilon. \quad (A124)$$

We employ now the "gentle measurement lemma" [40–42], saying that for all positive semidefinite operators $\sigma$, and $0 \leqslant H \leqslant 1$, one has

$$\|\sigma - \sqrt{H}\sigma\sqrt{H}\| \leqslant 2\sqrt{\mathrm{Tr}(\sigma)}\sqrt{\mathrm{Tr}(\sigma(I - H))}. \quad (A125)$$

Since $\Pi$ is a projector, and $\rho^\Gamma$ is normalized, from Eqs. (A122), (A123), and (A125), we find

$$\|\rho^\Gamma - \rho^\Gamma_\Pi\| \leqslant 2\sqrt{1 - \mathrm{Tr}(\Pi \rho^\Gamma \Pi)} \leqslant 2\sqrt{\epsilon}, \quad (A126)$$

where we used cyclic property of the trace. Using the triangle inequality twice, the fact that $\|\rho^\Gamma_\Pi\| \equiv \|\mathrm{diag}(\rho^\Gamma_\Pi)\|$, and

inequalities in Eqs. (A121), (A124), and (A126):

$$\left\|\rho^\Gamma - \frac{\mathrm{diag}(\rho^\Gamma_\Pi)}{\|\rho^\Gamma_\Pi\|}\right\| \leqslant \|\rho^\Gamma - \mathrm{diag}(\rho^\Gamma_\Pi)\| \quad (A127)$$

$$+ \left\|\mathrm{diag}(\rho^\Gamma_\Pi) - \frac{\mathrm{diag}(\rho^\Gamma_\Pi)}{\|\rho^\Gamma_\Pi\|}\right\| \quad (A128)$$

$$= \|\rho^\Gamma - \rho^\Gamma_\Pi + (\rho^\Gamma_\Pi - \mathrm{diag}(\rho^\Gamma_\Pi))\| + (1 - \|\rho^\Gamma_\Pi\|) \quad (A129)$$

$$\leqslant \|\rho^\Gamma - \rho^\Gamma_\Pi\| + \|\rho^\Gamma_\Pi - \mathrm{diag}(\rho^\Gamma_\Pi)\| + \epsilon \quad (A130)$$

$$\leqslant 2\sqrt{\epsilon} + \epsilon + \epsilon = 2(\sqrt{\epsilon} + \epsilon). \quad (A131)$$

This upper bound is tighter than the corresponding one for a pbit from proposition 1 due to application of corollary 3.

From the Refs. [15,22], the two-way repeater rate is upper bounded in the following way:

$$R^\leftrightarrow(\rho) \leqslant K_D(\rho^\Gamma) \leqslant E_r(\rho^\Gamma). \quad (A132)$$

While employing asymptotic continuity of the relative entropy of entanglement $E_r$ [43,44], we obtain

$$\left|E_r(\rho^\Gamma) - E_r\left(\frac{\mathrm{diag}(\rho^\Gamma_\Pi)}{\|\rho^\Gamma_\Pi\|}\right)\right| \leqslant \xi \log_2 \dim_H(\rho^\Gamma)$$

$$+ (1 + \xi)h\left(\frac{\xi}{1 + \xi}\right) \quad (A133)$$

$$\Rightarrow E_r(\rho^\Gamma) \leqslant E_r\left(\frac{\mathrm{diag}(\rho^\Gamma_\Pi)}{\|\rho^\Gamma_\Pi\|}\right)$$

$$+ \xi \log_2 \dim_H(\rho^\Gamma) + (1 + \xi)h\left(\frac{\xi}{1 + \xi}\right), \quad (A134)$$

where $\xi = 2(\sqrt{\epsilon} + \epsilon)$. Since dimension of a matrix is invariant under the partial transpose we have now:

$$R^\leftrightarrow(\rho) \leqslant E_r\left(\frac{\mathrm{diag}(\rho^\Gamma_\Pi)}{\|\rho^\Gamma_\Pi\|}\right) + \xi \log_2 \dim_H(\rho) \quad (A135)$$

$$+ (1 + \xi)h\left(\frac{\xi}{1 + \xi}\right). \quad (A136)$$

Blocks of $\mathrm{diag}(\rho^\Gamma)$ are separable from assumption. Since nonzero blocks $\mathrm{diag}(\rho^\Gamma_\Pi)$ are identical to corresponding blocks of $\mathrm{diag}(\rho^\Gamma)$ they are also separable. This implies that the relative entropy of entanglement of $\frac{\mathrm{diag}(\rho^\Gamma_\Pi)}{\|\rho^\Gamma_\Pi\|}$, from its definition reads 0, hence

$$R^\leftrightarrow(\rho) \leqslant 2(\sqrt{\epsilon} + \epsilon)\dim_H(\rho) \quad (A137)$$

$$+ (1 + 2\sqrt{\epsilon} + 2\epsilon)h\left(\frac{\sqrt{\epsilon} + \epsilon}{\frac{1}{2} + \sqrt{\epsilon} + \epsilon}\right). \quad (A138)$$

∎

*Proof of theorem 5.* We work under assumption that $\frac{d_k - 1}{d_s + d_k(d_k - 1)} \leqslant \|\rho - \gamma_{\langle d_k,d_s \rangle}\| \leqslant \epsilon < \frac{1}{d_k}$.

The first step is to upper bound key rate with relative entropy (see Ref. [22]).

$$K_D(\rho) \leqslant E_r(\rho). \tag{A139}$$

Then we make use of asymptotic continuity of quantum relative entropy [43,44].

$$\left| E_r(\rho) - E_r(\gamma_{\langle d_k, d_s \rangle}) \right| \leqslant \frac{\epsilon}{2} \log_2 \dim_H(\rho) + \left(1 + \frac{\epsilon}{2}\right) h\left(\frac{\frac{\epsilon}{2}}{1 + \frac{\epsilon}{2}}\right) \tag{A140}$$

$$\Rightarrow E_r(\rho) \leqslant E_r(\gamma_{\langle d_k, d_s \rangle}) + \frac{\epsilon}{2} \log_2 \dim_H(\rho) + \left(1 + \frac{\epsilon}{2}\right) h\left(\frac{\frac{\epsilon}{2}}{1 + \frac{\epsilon}{2}}\right). \tag{A141}$$

Since $E_r(\gamma_{\langle d_k, d_s \rangle}) \leqslant \log_2 d_k$ [22], by combining Eqs. (A139) and (A141), we have:

$$K_D(\rho) \leqslant \log_2 d_k + \frac{\epsilon}{2} \log_2 \dim_H \rho + \left(1 + \frac{\epsilon}{2}\right) h\left(\frac{\frac{\epsilon}{2}}{1 + \frac{\epsilon}{2}}\right). \tag{A142}$$

From Lemma 2, we know that $d_s \geqslant (\frac{d_k-1}{\epsilon})(1 - \epsilon d_k)$. We assume RHS to be positive, which together with the initial condition yields:

$$\log_2 d_s \geqslant \log_2 \left(\frac{d_k - 1}{\epsilon}\right) + \log_2(1 - \epsilon d_k). \tag{A143}$$

The overhead of the scheme is then lower bounded as follows:

$V(\rho)$

$$= M(\rho)\left(1 - \frac{K_D(\rho)}{\log_2 \dim_H(\rho)}\right) \tag{A144}$$

$$\geqslant M(\rho)\left(1 - \frac{\log_2 d_k + \frac{\epsilon}{2} \log_2 \dim_H(\rho) + \left(1 + \frac{\epsilon}{2}\right) h\left(\frac{\frac{\epsilon}{2}}{1+\frac{\epsilon}{2}}\right)}{\log_2 \dim_H(\rho)}\right) \tag{A145}$$

$$= M(\rho)\left(1 - \frac{\log_2 d_k + \left(1 + \frac{\epsilon}{2}\right) h\left(\frac{\frac{\epsilon}{2}}{1+\frac{\epsilon}{2}}\right)}{\log_2 d_k + \log_2 d_s} - \frac{\epsilon}{2}\right) \tag{A146}$$

$$\geqslant M(\rho)\left(1 - \frac{\log_2 d_k + \left(1 + \frac{\epsilon}{2}\right) h\left(\frac{\frac{\epsilon}{2}}{1+\frac{\epsilon}{2}}\right)}{\log_2 d_k + \log_2 \left(\frac{d_k-1}{\epsilon}\right) + \log_2(1 - \epsilon d_k)} - \frac{\epsilon}{2}\right). \tag{A147}$$

Now we have to find an appropriate lower bound on $K_D$. Following arguments of Ref. [45] the operation of privacy squeezing does not increase the trace distance $||\rho^{\mathrm{ps}} - \gamma^{\mathrm{ps}}_{\langle d_k, d_s \rangle}|| \leqslant \epsilon$, in a similar manner the key rate $K_D^{\rightarrow}(\rho^{\mathrm{ps}}) \leqslant K_D^{\rightarrow}(\rho)$. Moreover after this operation private state (strictly irreducible in that case) turns into maximally entangled state of dimension $d_k^2$.

$$K_D^{\rightarrow}(\rho^{\mathrm{ps}}) \leqslant K_D^{\rightarrow}(\rho) \leqslant K_D(\rho). \tag{A148}$$

On the other hand due to results in Ref. [46] and the fact that $K_D^{\rightarrow}(\rho^{\mathrm{ps}}) = \log_2 \dim_H(\rho^{\mathrm{ps}})$ both one-way and two-way keys are lower bounded

$$\log_2 d_k - 8\epsilon \log_2 \dim_H\left(\gamma^{\mathrm{ps}}_{\langle d_k, d_s \rangle}\right) - 4h(\epsilon) \leqslant K_D^{\rightarrow}(\rho^{\mathrm{ps}}). \tag{A149}$$

From Eqs. (A148) and (A149), and the fact $\dim_H(\gamma^{\mathrm{ps}}_{\langle d_k, d_s \rangle}) = d_k$, we obtain

$$K_D(\rho) \geqslant \eta := \log_2 d_k - 8\epsilon \log_2 d_k - 4h(\epsilon). \tag{A150}$$

Form proposition 2, the key repeater rate is upper bounded with

$$R^{\leftrightarrow}(\rho) \leqslant \theta := 2(\sqrt{\epsilon} + \epsilon) \dim_H(\rho) \tag{A151}$$

$$+ (1 + 2\sqrt{\epsilon} + 2\epsilon) h\left(\frac{\sqrt{\epsilon} + \epsilon}{\frac{1}{2} + \sqrt{\epsilon} + \epsilon}\right). \tag{A152}$$

∎